\documentclass[a4paper,12pt]{article}
\usepackage{amsmath}
\usepackage{amssymb}
\usepackage{amsfonts}
\usepackage{bbm} 
\usepackage{fleqn} 
\usepackage[footnotesize]{caption}
\usepackage{graphicx} 
\usepackage{braket} 
\usepackage{mathrsfs}
\usepackage[center,footnotesize,hang]{subfigure}

\newcommand{\CenterEps}[2][1]{\ensuremath{\vcenter{\hbox{\includegraphics[scale=#1]{#2.eps}}}}}
\newcommand{\SuperField}[1]{\hat{#1}}
  
\def\Ng{j} 
\def\Nf{i}

\def\I{\mathrm{i}}
\def\SU{\text{SU}}
\def\SO{\text{SO}}
\def\U{\text{U}}

\def\SingletH{\SuperField{\theta}}
\def\SingletHB{{\theta}}

\def\SingletHi{(\SuperField{\theta}_I){}}

\def\SingletHaB{({\theta}_1){}}
\def\SingletHbB{({\theta}_2){}}
\def\SingletHcB{({\theta}_3){}}

\def\SingletHiB{({\theta}_I){}}
\def\SingletHjB{({\theta}_J){}}

\def\SingletHe{\SuperField{\chi}}

\def\MSingletHa{M_{N1}{}}
\def\MSingletHb{M_{N2}{}}
\def\MSingletHc{M_{N3}{}}
\def\MSingletHi{M_{N I}{}}
\def\MSingletHj{M_{N J}{}}
\def\MSingletHk{M_{N K}{}}

\def\SingletHei{(\SuperField{\chi}_I){}}

\def\SingletHeaB{({\chi}_1){}}
\def\SingletHebB{({\chi}_2){}}
\def\SingletHecB{({\chi}_3){}}

\def\MSingletHea{M_{E1}{}}
\def\MSingletHeb{M_{E2}{}}
\def\MSingletHec{M_{E3}{}}
\def\MSingletHei{M_{E I}{}}
\def\MSingletHej{M_{E J}{}}
\def\MSingletHek{M_{E K}{}}

\def\epsa{{\varepsilon}_1}
\def\epsb{{\varepsilon}_2}
\def\epsc{{\varepsilon}_3}

\def\da{{\delta}_1}
\def\db{{\delta}_2}
\def\dc{{\delta}_3}

\def\epsaP{{\varepsilon}'_1}
\def\epsbP{{\varepsilon}'_2}
\def\epscP{{\varepsilon}'_3}

\def\daP{{\delta}'_1}
\def\dbP{{\delta}'_2}
\def\dcP{{\delta}'_3}

\def\<{\left\langle}
\def\>{\right\rangle}
\def\ChargeC{\mathrm{C}} 
\def\chargec{\mathrm{C}}

\DeclareMathOperator{\Tr}{Tr}

\allowdisplaybreaks[2]

\begin{document}

\bibliographystyle{OurBibTeX}

\begin{titlepage}

 \vspace*{-15mm}
\begin{flushright}
SHEP/0403
\end{flushright}
\vspace*{5mm}

\begin{center}
{
\sffamily\LARGE 
From Hierarchical to Partially Degenerate Neutrinos  \\[1mm] 
via  Type II Upgrade of Type I See-Saw Models
}
\\[12mm]
S. Antusch\footnote{E-mail: \texttt{santusch@hep.phys.soton.ac.uk}},
S. F. King\footnote{E-mail: \texttt{sfk@hep.phys.soton.ac.uk}}
\\[1mm]
{\small\it
Department of Physics and Astronomy,
University of Southampton,\\
Southampton, SO17 1BJ, U.K.
}
\end{center}
\vspace*{1.00cm}

\begin{abstract}

\noindent We propose a type II upgrade of type I see-saw models leading 
to new classes of models where partially degenerate neutrinos are 
as natural as hierarchical ones. 
The additional type II contribution to the neutrino mass matrix, 
which determines the neutrino mass scale, is forced to be proportional 
to the unit matrix by a SO(3) flavour symmetry.  
The type I see-saw part of the neutrino mass matrix, which controls the  
mass squared differences and mixing angles, 
may be governed by sequential right-handed neutrino dominance and 
a natural alignment for the SO(3)-breaking vacuum. 
We focus on classes of models with bi-large mixing originating from the neutrino 
mass matrix although we also briefly discuss other classes of models where large 
mixing stems from the charged lepton mass matrix.
We study renormalization group corrections to the neutrino mass squared
differences and mixings and find that the low energy values do not depend
sensitively on the high energy values for partially degenerate neutrinos 
with a mass scale up to about 0.15 eV.
Our scenario predicts the effective mass for neutrinoless double beta decay 
to be approximately equal to the neutrino mass scale and therefore 
neutrinoless double beta decay will be observable if the neutrino mass 
spectrum is partially degenerate. We also find that all observable CP phases 
as well as $\theta_{13}$ become small as the neutrino mass scale increases. 
\end{abstract}

\end{titlepage}
\newpage
\setcounter{footnote}{0}

\section{Introduction}

The observation of flavour conversions of neutrinos together with their
interpretation by neutrino oscillations has brought crucial new information 
about fermion masses and mixings. Neutrinos are massive, with very small 
measured mass squared differences, and contrary to the quark sector, 
there is large flavour mixing among the leptons. 
The present 3$\sigma$ ranges 
for the parameters are 
$\theta_{12} \in [27.6^\circ,36.3^\circ]$, 
$\Delta m^2_\mathrm{sol}:= m_2^2 - m_1^2 \in 
[5.4\cdot 10^{-5}\:\mbox{eV}^2,9.5\cdot 10^{-5}\:\mbox{eV}^2]$  ,
 $\theta_{23} \in [32.2^\circ,51.4^\circ]$,  
$|\Delta m^2_\mathrm{atm}|:= |m_3^2 - m_1^2| \in 
[1.4\cdot 10^{-3}\:\mbox{eV}^2,3.7\cdot 10^{-3}\:\mbox{eV}^2]$ 
 and for $\theta_{13}$  the current upper bound which mainly stems from the
 CHOOZ data \cite{Apollonio:1999ae} is 
$\theta_{13} \lesssim 15^\circ$. 
The values have been taken from the global analysis \cite{Maltoni:2003da}  
which includes the
the Super-Kamiokande atmospheric data
 \cite{Toshito:2001dk}, the KamLAND results \cite{Eguchi:2002dm} and 
 recent SNO salt results \cite{Ahmed:2003kj}. 

One of the most interesting missing piece of information is the 
 neutrino mass scale.  
 At present, the most stringent bounds are 
 $m_i < 0.23$~eV  
 from WMAP 
 \cite{Spergel:2003cb} and $\Braket{m_\nu}\lesssim 0.35\,\mathrm{eV}$,
 with some uncertainty  
 due to nuclear matrix elements,
  from neutrino-less
 double beta ($00\nu\beta$) decay experiments 
 \cite{Klapdor-Kleingrothaus:2000sn,Aalseth:2002rf}. 
 The latter search for 
 an effective mass defined by 
 $\Braket{m_\nu}= \left| \sum_i (U_\mathrm{MNS})_{1i}^2 \, m_i \right| 
$ and are exclusively sensitive to Majorana masses. 	
Future experiments which are under consideration at present might increase the 
sensitivity to $\Braket{m_\nu}$ by more than an order of magnitude. 
The neutrino mass spectrum for the mass range accessible to this next round of 
$0\nu\beta\beta $ decay searches shows at least a partial degeneracy. 
It is therefore interesting to investigate theoretical scenarios which could 
account for such neutrino mass schemes.

The most promising scenarios for giving masses to neutrinos
use a version of the see-saw mechanism 
\cite{Yanagida:1980,Glashow:1979vf,Gell-Mann:1980vs,Mohapatra:1980ia}, which provides a convincing
explanation for their smallness.
Models for strongly degenerate neutrinos have been considered e.g.~in 
\cite{Caldwell:1993kn,Bamert:1994vc,Lee:1994qx,Ioannisian:1994nx,Joshipura:1994jy,Joshipura:1995ax,Ghosal:1997vs,Carone:1998bb,Ma:1998db,Ray:1998eq,Lazarides:1998jt,Wetterich:1998vh,Wu:1998if,Ghosal:1999jb,Barbieri:1999km,Wu:1999mu,Wu:1999yz,Ma:2001dn,Babu:2002dz,Patgiri:2003ah,Mohapatra:2003tw}.
 In most of them, the degeneracy is achieved 
by discrete symmetries. The non-Abelian flavour symmetry group $\SO(3)$ in 
connection with degenerate neutrinos has e.g.~been considered in 
 \cite{Wu:1998if,Carone:1998bb,Ma:1998db,Wetterich:1998vh,Barbieri:1999km}. 
It has turned out that using the type I see-saw mechanism 
 in order to explain
the smallness of neutrino masses, it seems to be 
difficult, if not impossible, to obtain a nearly degenerate neutrino mass 
spectrum in a natural way.  
In order to find natural explanations for partially degenerate neutrino masses, 
it is therefore promising to consider the type II  
see-saw mechanism  
(see e.g.~\cite{Lazarides:1980nt,Mohapatra:1981yp,Wetterich:1981bx,Ma:1998dx}),  
as has been argued for example in \cite{Caldwell:1993kn,Joshipura:1995ax}. 

In this work, we propose a type II upgrade of type I see-saw models leading 
to new classes of models where partially degenerate neutrinos are 
as natural as hierarchical ones. 
This is achieved by a SO(3) flavour symmetry, which 
forces the additional type II mass term to be proportional 
to the unit matrix in leading order. The addition of a type II unit matrix 
contribution to the type I neutrino 
mass matrix with a particular phase structure turns out to enable the neutrino  
mass scale to be increased almost arbitrarily, while leaving the mixing angles 
approximately unchanged. In this approach  
the type I see-saw part of the neutrino mass matrix, which controls the  
mass squared differences and mixing angles, 
may be governed by sequential right-handed neutrino dominance and 
a natural alignment for the SO(3)-breaking vacuum. 
We focus on classes of models with bi-large mixing originating from the neutrino 
mass matrix although we also briefly discuss other classes of models where large 
mixing stems from the charged lepton mass matrix. 
We also study renormalization group corrections to the mass squared 
differences and mixings and find that the low energy values do not depend 
sensitively on the high energy values for partially degenerate neutrinos  
with a mass scale up to about 0.15 eV. 
This framework predicts the effective mass for neutrinoless double beta decay  
to be approximately equal to the neutrino mass scale and therefore  
neutrinoless double beta decay will be observable if the neutrino mass  
spectrum is partially degenerate. We also find that all observable CP phases  
become small as the neutrino mass scale increases. 

The layout of the paper is as follows: 
After giving a motivation in section 2, we outline how naturally small 
neutrino masses could emerge from 
type I and 
type II see-saw mechanisms in section 3. In section 4 we discuss the 
type II see-saw scenario with spontaneously broken SO(3) flavour symmetry 
and analyze the consequences for the ingredients of the 
type II see-saw formula. In section 5 we consider a real alignment 
mechanism for the SO(3)-breaking vacuum expectation values (vevs) and show 
how it leads to classes of models where the observed bi-large 
neutrino mixing can naturally be obtained. 
In section 6 we focus on models with 
sequential right-handed neutrino dominance \cite{King:1999mb,King:2002nf} 
for the type I part of the 
neutrino mass matrix.  
We extract the mixing angles, masses and CP phases analytically and discuss the 
predictions for these parameters from our scenario. In section 7 we study  
renormalization group corrections to the neutrino mass squared
differences and mixings. In section 8 we analyze the predictions for 
the effective mass for neutrinoless
double beta decay. Section 9 contains a  
discussion and our conclusions.   

\section{Motivation}
\begin{figure}[h]
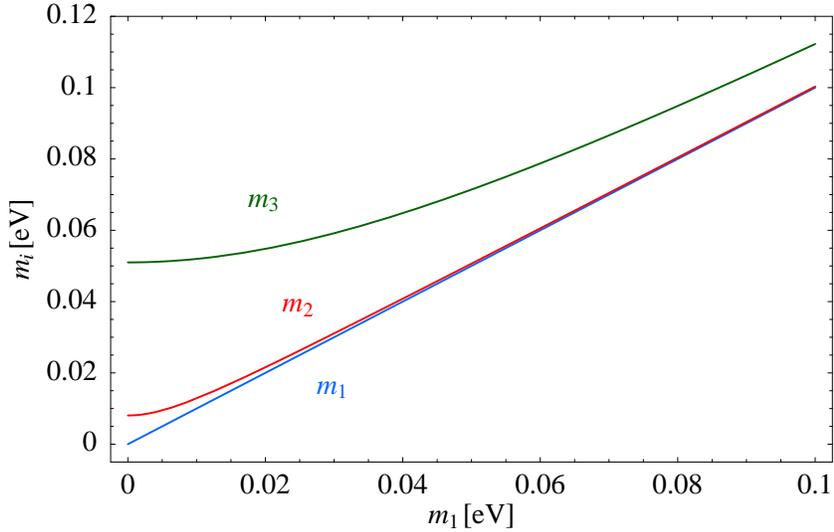

 \begin{center}
\CenterEps[0.9]{NeutrinoMasses}
 \end{center}
 \caption{\label{fig:NuMasses}
 Neutrino masses as a function of the mass of the lightest
 neutrino in a normal mass scheme, where $m_3 > m_2 > m_1$.  
 }
\end{figure}

For a given mass of the lightest neutrino and e.g.~a normal scheme for the
neutrino masses $m_1 < m_2 < m_3$, the remaining two masses can be calculated
from the requirement that the experimentally measured mass squared differences
are produced.   
Figure \ref{fig:NuMasses} shows the neutrino mass eigenvalues as a function 
of the mass of the lightest neutrino. For a lightest neutrino 
 heavier than about $0.02$ eV, the mass eigenvalues of the neutrinos are of the
 same order.  
We will refer to the neutrinos as partially degenerate, if 
the mass of the lightest neutrino is roughly in the range $[0.02,0.15]$ eV, 
where $m_1$ and $m_2$ are nearly degenerate. 
This is below what is usually called quasi-degenerate where the masses of all
neutrinos are approximately degenerate.    
This mass range is particularly interesting, since it is not disfavored by
unnaturally large radiative corrections and it might be accessible to 
future $0\nu\beta\beta $ decay searches. 

\newpage

\vspace*{5mm}

\noindent With three left-handed neutrinos contained in the lepton doublets and 
three right-handed neutrinos which are singlets under 
$G_{321}:=
\SU (3)_\mathrm{C}\times \SU (2)_\mathrm{L} \times \U (1)_\mathrm{Y}$, 
the general neutrino mass matrix is given by
\begin{eqnarray}
\mathscr{L}_{M_\nu} &=&  - \frac{1}{2}\,
\left( \begin{array}{c} \overline{\nu_{\mathrm{L}}^{f}}  \\[1mm]
  \overline{\nu^{\ChargeC i}_\mathrm{R}}   \end{array}   \right)^T
\left( \begin{array}{cc} 
(m^{\mathrm{II}}_{\mathrm{LL}})_{fg} \vphantom{\nu_{\mathrm{L}}^{\ChargeC g}}&
(m_{\mathrm{LR}})_{fj}\\[1mm]
 (m_{\mathrm{LR}}^T)_{ig}\vphantom{\nu_{\mathrm{L}}^{\ChargeC g}}&
 (M_{\mathrm{RR}})_{ij}
    \end{array}   \right)
  \,
 \left( \begin{array}{c} 
\nu_{\mathrm{L}}^{\ChargeC g}  \\[1mm]
  {\nu^{ j}_\mathrm{R}}   \end{array}   \right)+\text{h.c.}\; .
\end{eqnarray}
Under the assumption that the mass eigenvalues $M_{\mathrm{R}i}$ of
$M_{\mathrm{RR}}$ are very large compared to the components of  $m^{\mathrm{II}}_{\mathrm{LL}}$
and $m_{\mathrm{LR}}$, the mass matrix can approximately be diagonalized yielding  
\begin{eqnarray}
\mathscr{L}_{M_\nu} &\approx&  - \frac{1}{2}\,
\left( \begin{array}{c} \overline{\nu{}'{}_{\mathrm{L}}^{f}}  \\[1mm]
  \overline{\nu{}'{}^{\ChargeC i}_\mathrm{R}}   \end{array}   \right)^T
\left( \begin{array}{cc} 
(m^{\nu}_{\mathrm{LL}})_{fg} \vphantom{\nu_{\mathrm{L}}^{\ChargeC g}}&
0\\[1mm]
 0\vphantom{\nu_{\mathrm{L}}^{\ChargeC g}}& (M_{\mathrm{RR}})_{ij}
    \end{array}   \right)
  \,
 \left( \begin{array}{c} 
\nu{}'{}_{\mathrm{L}}^{\ChargeC g}  \\[1mm]
  {\nu{}'{}^{ j}_\mathrm{R}}   \end{array}   \right)+\text{h.c.}\; ,
\end{eqnarray}
where, neglecting $\mathscr{O}(M_{\mathrm{R}i}^{-1})$-terms, 
$\nu{}{'}_{\mathrm{L}}^f\approx\nu_{\mathrm{L}}^f$ and 
$\nu{}{'}{}^{\ChargeC i}_\mathrm{R}\approx \nu^{\ChargeC i}_\mathrm{R}$.
The Majorana mass matrix $m^\nu_{\mathrm{LL}}$ 
for the light left-handed neutrinos is given by
\begin{eqnarray}\label{eq:TypIIMassMatrix}
m^\nu_{\mathrm{LL}} \approx 
m^{\mathrm{II}}_{\mathrm{LL}} + m^{\mathrm{I}}_{\mathrm{LL}}
\end{eqnarray} 
with 
\begin{eqnarray}
m^{\mathrm{I}}_{\mathrm{LL}} := - m_{\mathrm{LR}}
\,M^{-1}_{\mathrm{RR}}\,m^T_{\mathrm{LR}}\; .
\end{eqnarray}
The suppression by $M_{\mathrm{RR}}^{-1}$ provides a natural explanation 
for the smallness of neutrino masses from $m^{\mathrm{I}}_{\mathrm{LL}}$. This is 
referred to as the type I see-saw mechanism 
\cite{Yanagida:1980,Glashow:1979vf,Gell-Mann:1980vs,Mohapatra:1980ia}. 
As we will sketch in section \ref{sec:SeeSaw}, the direct mass term  
$m^{\mathrm{II}}_{\mathrm{LL}}$ can also provide a naturally small contribution to the 
light neutrino masses if it stems e.g.~from a see-saw suppressed induced vev. 
We will refer to the general case,   
where both possibilities are allowed, as the II see-saw mechanism 
\cite{Lazarides:1980nt,Mohapatra:1981yp,Wetterich:1981bx,Ma:1998dx}.

In type I see-saw models, it seems to be 
difficult, if not impossible, to obtain a partially degenerate or
quasi-degenerate neutrino mass 
spectrum in a natural way, whereas hierarchical masses seem to be natural.   
For a review on neutrino mass models, see e.g.~\cite{King:2003jb}. 
The direct mass term in type II models on the other hand has the potential to 
provide a natural way for generating neutrino masses with a partial degeneracy. 
Imagine for example that by symmetry, the direct mass term is forced to be
proportional to the unit matrix in flavour space, 
$m^{\mathrm{II}}_{\mathrm{LL}} = m^{\mathrm{II}} \,\mathbbm{1}$.  
 We will realize such a direct mass term,  
 which gives a common mass to all the neutrinos,   
 in section \ref{sec:SO(3)} via 
 SO(3) flavour symmetry.
The type II formula of equation (\ref{eq:TypIIMassMatrix}) is then realized by
\begin{eqnarray}
m^\nu_{\mathrm{LL}} \approx m^{\mathrm{II}}\, 
\left(\begin{array}{ccc}
1&0&0\\
0&1&0\\
0&0&1
\end{array}\right)
 +
 \left(\begin{array}{ccc}
(m^{\mathrm{I}}_{\mathrm{LL}})_{11} &(m^{\mathrm{I}}_{\mathrm{LL}})_{12} & (m^{\mathrm{I}}_{\mathrm{LL}})_{13}\\
(m^{\mathrm{I}}_{\mathrm{LL}})_{21} &(m^{\mathrm{I}}_{\mathrm{LL}})_{22} &(m^{\mathrm{I}}_{\mathrm{LL}})_{23}\\
(m^{\mathrm{I}}_{\mathrm{LL}})_{31} &(m^{\mathrm{I}}_{\mathrm{LL}})_{32} &(m^{\mathrm{I}}_{\mathrm{LL}})_{33}
\end{array}\right)
\end{eqnarray}  
 and the direct mass term $m^{\mathrm{II}}\, \mathbbm{1}$ naturally allows for partially 
 degenerate neutrinos. 
 The neutrino mass splittings and mixing angles in this scenario 
 are mainly controlled by 
 the type I see-saw contribution $m_{\mathrm{LL}}^{\mathrm{I}} $. 
 To analyze the effect of the type II contribution $m_{\mathrm{LL}}^{\mathrm{II}}$, 
 we consider the diagonalization of $m_{\mathrm{LL}}^{\mathrm{I}}$ 
 by a unitary transformation 
 $(m_{\mathrm{LL}}^{\mathrm{I}})_{\mathrm{diag}} = V 
  m_{\mathrm{LL}}^{\mathrm{I}} V^T $. 
 If we assume for the moment that the type I see-saw
 mass matrix $m_{\mathrm{LL}}^{\mathrm{I}}$ is real, which implies that $V$ is an 
 orthogonal matrix, we obtain  
  \begin{eqnarray}\label{eq:TypIISeeSawFormulaUnitMatrix}
 (m^\nu_{\mathrm{LL}})_{\mathrm{diag}} \;= \; 
 m^{\mathrm{II}}\, V V^T +  V  m_{\mathrm{LL}}^{\mathrm{I}} V^T
\;=\; m^{\mathrm{II}}\, \mathbbm{1} + (m_{\mathrm{LL}}^{\mathrm{I}})_{\mathrm{diag}}
\; . 
 \end{eqnarray}
The additional direct mass term leaves the predictions for 
the mixings from the type I see-saw contribution unchanged in this case. 
 This allows to transform many   
type I see-saw models for hierarchical neutrino 
masses into type II see-saw models for partially degenerate or
quasi-degenerate neutrino masses while maintaining the predictions for the 
mixing angles. 
Obviously, in the general complex case, 
it is no longer that simple since for a unitary matrix $V V^T
\not= \mathbbm{1}$ and the phases will have impact on the predictions for the
mixings. 
We will return to this issue in section \ref{sec:Typw2SeqRhdNuDom}  
where we will see that e.g.~in some classes of models with sequential 
right-handed neutrino dominance (RHND) \cite{King:1999mb,King:2002nf} for the type I contribution to the neutrino 
mass matrix, the known techniques and mechanisms for explaining
the bi-large lepton mixings 
can be directly applied also in the presence of 
CP phases.

\section{Type I and Type II See-Saw Mechanisms}\label{sec:SeeSaw} 
\begin{figure}
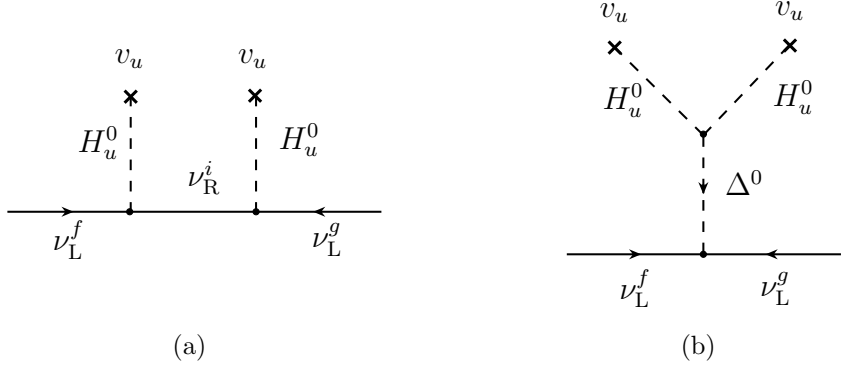

\begin{center}
  \subfigure[]{$
  \CenterEps[1]{typeIseesaw}\vphantom{\CenterEps[1]{typeIIseesaw}}
  $}  \hfil
  \subfigure[]{$
  \CenterEps[1]{typeIIseesaw}
  $} 
 \caption{\label{fig:TypeIIDiagrams}
Diagrams leading to neutrino masses in the type II  
see-saw scenario. 
Diagram (a) shows the contribution from the exchange of a heavy right-handed
neutrino as in the type I see-saw mechanism. Diagram (b) illustrates the 
 contribution from an induced vev of 
the triplet $\Delta$. At low energy, they can be viewed as contributions to
the effective neutrino mass operator from  
integrating out the heavy fields $\nu_{\mathrm{R}}^i$ and $\Delta^0$,  
respectively.  
 } 
\end{center}
\end{figure}

We now outline how naturally small neutrino masses could emerge from 
minimal realizations of type I and type II see-saw mechanisms. 
Let us consider a minimal extension of the MSSM in order to allow for 
 neutrino masses. We discuss the supersymmetric case 
here since the
modifications in oder to obtain the non-supersymmetric version 
are straightforward.
Dirac masses for the neutrinos can be achieved by adding chiral superfields 
$\SuperField{\nu}^{\ChargeC \Nf}$ $(\Nf\in \{1,2,3\})$   
in the representation $(\boldsymbol{1},\boldsymbol{1},0)$ of 
$G_{321}$ to the
particle content, 
which contain the right-handed neutrinos $\nu^{\Nf}_\mathrm{R}$ as  
fermionic components. 
If not protected by symmetry, these singlets are expected to
obtain large masses $M_{\mathrm{R}i}$, associated with the scale of lepton number
breaking. Lepton masses now arise from the superpotential
\begin{eqnarray}\label{eq:SuperpotentialInTheMSSM}
 \mathcal{W}_{\ell} & = &
 - \,(Y_e)_{gf}(\SuperField{L}^{g} \cdot 
 	\,\SuperField{H}_d )\,\SuperField{e}^{\chargec f}
 +(Y_\nu)_{f \Ng}(\SuperField{L}^{f}\cdot
 \SuperField{H}_u)\,  \SuperField{\nu}^{\chargec \Ng}
 + \frac{1}{2} \SuperField{\nu}^{\chargec \Nf} (M_\mathrm{RR})_{\Nf \Ng}
 \SuperField{\nu}^{\chargec \Ng}
 \; ,	
\end{eqnarray}
where the dot indicates the $\SU (2)_\mathrm{L}$ invariant product, 
i.e.~$(\SuperField{L}^{\Ng}\cdot \SuperField{H}_u) := 
\SuperField{L}_a^{\Ng}(i\tau_2)^{ab} (\SuperField{H}_u)_b$ with $\tau_A$  
$(A\in \{1,2,3\})$ being
the Pauli matrices. 
The superfields $\SuperField{H}_u$ and $\SuperField{H}_d$ contain the Higgs 
fields which also give masses to the up-type and down-type quarks respectively. 
$\SuperField{L}$ contains the lepton doublets and 
 $\SuperField{e}^{\chargec}$ the charged leptons. 
This yields naturally small neutrino masses 
by the type I see-saw relation 
\begin{eqnarray}\label{eq:TypISeeSawFormula}
m^{\mathrm{I}}_\mathrm{LL} = - \,  v^2_u\, 
Y_\nu \, M_\mathrm{RR}^{-1}\, Y_\nu^T \; .
\end{eqnarray} 
In many cases however, 
the left-handed neutrinos may also obtain a naturally small
 direct  mass term. 
This happens for example if $\SU (2)_\mathrm{L}$-triplet Higgs superfields 
$\SuperField{\Delta}$ 
and $\bar{\SuperField{\Delta}}$ are added to the particle spectrum which have
weak hypercharge $q_\mathrm{Y}=+1$ and $q_\mathrm{Y}=-1$, respectively. 
The representations of the chiral superfields involved in this minimal setup
are given in table \ref{tab:QuantumNumbersOfNuMSSM}.
\begin{table}
\begin{center}
\begin{tabular}{l|ccccccc}
	$\vphantom{\sqrt{\big|}}$Field &
		\(\SuperField{H}_d\) & \(\SuperField{H}_u\) 
		& \(\SuperField{\Delta}\)&
		\(\bar{\SuperField{\Delta}}\)
		& \(\SuperField{L}^{f}\) & \({\SuperField{e}}^{\ChargeC f}\) 
			& \(\SuperField{\nu}^{\ChargeC f}\) 
			 \\
	\hline
	\(\vphantom{\sqrt{\big|}^C}\mathrm{SU}(3)_\mathrm{C}\)	
		& \(\boldsymbol{1}\) & \(\boldsymbol{1}\)
		& \(\boldsymbol{1}\) & \(\boldsymbol{1}\) 
		& \(\boldsymbol{1}\) & \(\boldsymbol{1}\)& \(\boldsymbol{1}\)
		\\
	\(\vphantom{\sqrt{\big|}}\mathrm{SU}(2)_\mathrm{L}\)	
		& \(\boldsymbol{2}\) & \(\boldsymbol{2}\)
		& \(\boldsymbol{3}\) & \(\boldsymbol{3}\)
		& \(\boldsymbol{2}\) & \(\boldsymbol{1}\) & \(\boldsymbol{1}\)
		\\
	\(\vphantom{\sqrt{\big|}} q_\mathrm{Y}\)
		& \(-\tfrac{1}{2}\) & \(+\tfrac{1}{2}\)
		& \(1\) & \(-1\)
		& \(-\tfrac{1}{2}\) & \(+1\) & \(0\)	 
\end{tabular}
\caption{\label{tab:QuantumNumbersOfNuMSSM}
Representations under $G_{321}$ of the chiral superfields involved in the generation of lepton
masses in the MSSM extended by three singlet superfields which contain the
right-handed neutrinos and by two $\mathrm{SU}(2)_\mathrm{L}$-triplets. }
\end{center}
\end{table}
\noindent Only the superfield $\SuperField{\Delta}$ contributes to the generation of 
fermion masses via the coupling 
\begin{eqnarray}\label{eq:SuperpotentialDelta}
 \mathcal{W}_\Delta & = &
 \frac{1}{2} \,(Y^\dagger_\Delta)_{fg} \,\SuperField{L}^T{}^f\,i\tau_2\, \SuperField{\Delta} 
 \,\SuperField{L}^g\; 
\end{eqnarray} 
to the lepton doublets. We have written the 
$\mathrm{SU}(2)_\mathrm{L}$-triplets 
as traceless $2\times2$-matrices 
\begin{eqnarray}
 \SuperField{\Delta}= 
\left(\begin{array}{cc}
 \SuperField{\Delta}^{+}& \SuperField{\Delta}^{++} \\
 \SuperField{\Delta}^{0}  &-\SuperField{\Delta}^{+} 
	 \end{array}\right)  \:\mbox{and}\;\;
\bar{\SuperField{\Delta}}= 
 \left(\begin{array}{cc}
 \bar{\SuperField{\Delta}}^{+}& \bar{\SuperField{\Delta}}^{++} \\
 \bar{\SuperField{\Delta}}^{0}  &-\bar{\SuperField{\Delta}}^{+} 
	 \end{array}\right)  .	 
\end{eqnarray}
A vev $v_\Delta=\<{\Delta}^0\>$ of the neutral component of the scalar field contained 
in $\SuperField{\Delta}$ gives a direct mass for the left-handed neutrinos. 
We choose $v_\Delta$ to be real and positive by a proper 
phase choice for $\SuperField{\Delta}$.  
In order to estimate the natural size of $v_\Delta$, we consider the 
 Higgs potential from the part 
\begin{eqnarray}
\mathcal{W}_{H} = M_\Delta \Tr (\SuperField{\Delta} \,\bar{\SuperField{\Delta}})
+ \lambda_u  \,
\SuperField{H}^T_u\,i\tau_2\, \bar{\SuperField{\Delta}}\,\SuperField{H}_u
 +\lambda_d\,
\SuperField{H}^T_d\,i\tau_2\, \SuperField{\Delta}\,\SuperField{H}_d
 + \mu \,(\SuperField{H}_d\cdot\SuperField{H}_u) 
\end{eqnarray} 
of the superpotential. 
Using the expansion of the superfields 
$
\SuperField{H}_u = H_u + \sqrt{2}\, \theta 
\Tilde{H}_u + 
\theta\theta\,\,F_{H_u} 
$ and 
$
\SuperField{\Delta} = \Delta + \sqrt{2}\, \theta 
\Tilde{\Delta} + 
\theta\theta\,\,F_{\Delta}  
$, 
 we see that the scalar potential from $|F_{\bar{\Delta}}|^2$ contains the 
terms
\begin{eqnarray}
\mathcal{V} =  M_\Delta \lambda_u\, H^T_u \,i\tau_2\, \Delta^* H_u 
+  M^2_\Delta  \Tr(\Delta^* \Delta) \;+\;\text{h.c.}\; .
\end{eqnarray}
After EW symmetry breaking, this results in a tadpole which forces $\Delta$
to get an induced vev of the order 
\begin{eqnarray}
v_\Delta \approx \frac{\lambda_u v_u^2}{M_\Delta}\; .
\end{eqnarray}
Analogously, the neutral component of the superfield 
$\bar{\SuperField{\Delta}}$ obtains a small vev as well, however it is not
relevant here since 
$\bar{\SuperField{\Delta}}$ 
does
not couple to the fermions. 
If $M_\Delta$ is large, say of the order of one of the eigenvalues of 
$M_\mathrm{RR}$, this leads to another naturally small 
contribution 
\begin{eqnarray}
m^{\mathrm{II}}_{\mathrm{LL}} = Y_{\Delta} v_\Delta
\end{eqnarray}
to the neutrino mass matrix, which is now given by 
the type II see-saw formula 
 \begin{eqnarray}\label{eq:TypIISeeSawFormula}
 m^\nu_\mathrm{LL}  = m^{\mathrm{II}}_{\mathrm{LL}} + m^{\mathrm{I}}_{\mathrm{LL}} =  Y_{\Delta} v_\Delta - v^2_u
\, Y_\nu \, M_\mathrm{RR}^{-1}\, Y_\nu^T \; .
 \end{eqnarray}
 The contributions to neutrino masses in 
the considered minimal type II scenario are illustrated in figure
\ref{fig:TypeIIDiagrams}. 
 Note that we do not require left-right symmetry 
throughout this paper. 
Additional triplets like the ones used here to provide a minimal example for 
 a type II see-saw mechanism typically appear in models with a left-right 
 symmetric particle content like minimal left-right 
symmetric models \cite{Pati:1974yy,Mohapatra:1975gc,Senjanovic:1975rk},  
Pati-Salam models \cite{Pati:1973uk} or $\SO (10)$-GUTs 
\cite{GeorgiSO10,Fritzsch:1975nn}.  
However, a naturally small direct vev can also originate from other sources like 
e.g.~from higher-dimensional operators and most of 
the results of this paper can also be applied to this case.

\section{Type II See-Saw with $\SO(3)$ Flavour Symmetry}\label{sec:SO(3)}
We now show how a contribution to the neutrino mass matrix proportional to the 
unit matrix can be achieved by a spontaneously broken 
$\SO(3)$ flavour symmetry. 
We will analyze the consequences of such a framework 
 for the masses and mixings in the lepton sector, where we 
 consider the case that $\SO(3)$ acts on the 
lepton doublets. This has impact on the ingredients of the 
 type II see-saw formula for the neutrino mass matrix 
 of equation (\ref{eq:TypIISeeSawFormula}) 
as well as for the mass matrix of the charged leptons.

\subsection{Structure of the Ingredients of the Type II See-Saw} 
In order to break $\SO(3)$ 
spontaneously, we introduce additional heavy $G_{321}$-singlet 
superfields 
$\SingletH_{I}$ ($I \in \{1,2,3\}$) which are flavour triplets 
and acquire vevs $\<\SingletHiB_{f}\>$. 
In order to associate each flavon field $\SingletH_{I}$ with one of the right-handed 
neutrino superfields $\SuperField{\nu}^{\ChargeC i}$, we introduce 
three discrete symmetries ${Z}^{I}_2$. The representations of the 
fields involved in the generation of the neutrino mass matrix 
are given in table \ref{tab:repsSO(3)}.

\begin{table}[h]
\begin{center}
\begin{tabular}{l|ccccccccc}
	$\vphantom{\sqrt{\big|}}$Field &
        \(\SuperField{L}\) & \(\SuperField{\Delta}\) & \(\SuperField{H}_u\) 
	& \(\SuperField{\nu}^{\ChargeC 1}\) & \(\SuperField{\nu}^{\ChargeC 2}\)& \(\SuperField{\nu}^{\ChargeC 3}\)
	& \(\SingletH_1\)& \(\SingletH_2\)& \(\SingletH_3\)
			\\
	\hline				
	\(\vphantom{\sqrt{\big|}} \SO(3)\)
	& \(\boldsymbol{3}\)& \(\boldsymbol{1}\)& \(\boldsymbol{1}\)
	& \(\boldsymbol{1}\) & \(\boldsymbol{1}\) & \(\boldsymbol{1}\) 
	& \(\boldsymbol{3}\) & \(\boldsymbol{3}\) & \(\boldsymbol{3}\)
		\\	
	\(\vphantom{\sqrt{\big|}}{Z}^1_2 \)
	& \(+\)& \(+\)& \(+\)
	& \(-\) & \(+\) & \(+\) 
	& \(-\) & \(+\) & \(+\)	
		\\
	\(\vphantom{\sqrt{\big|}}{Z}^2_2 \)
	& \(+\)& \(+\)& \(+\)
	& \(+\) & \(-\) & \(+\) 
	& \(+\) & \(-\) & \(+\)	
		\\
	\(\vphantom{\sqrt{\big|}}{Z}^3_2 \)
	& \(+\)& \(+\)& \(+\)
	& \(+\) & \(+\) & \(-\) 
	& \(+\) & \(+\) & \(-\)	
\end{tabular}	
\end{center}
\caption{
\label{tab:repsSO(3)} 
Representations of $\SO(3) 
\times \:({Z}_2)^3$ involved in the generation of neutrino masses. 
}
\end{table}	 

\noindent The renormalizable term for neutrino Yukawa couplings is forbidden by the 
$\SO(3)$ flavour symmetry. Dirac masses for the neutrinos
can however be generated by the higher-dimensional operators
\begin{eqnarray}\label{eq:effectiveYnu}
\mathcal{W}_{Y_\nu} &=& 
a_1 \,(\SuperField{L}^f\cdot \SuperField{H}_u)\, \SuperField{\nu}^{\chargec 1} \, 
 \frac{\<\SingletHaB_{f}\> }{\MSingletHa}+
 a_2\,( \SuperField{L}^f\cdot\SuperField{H}_u)\, \SuperField{\nu}^{\chargec 2} \, 
 \frac{\<\SingletHbB_{f}\> }{\MSingletHb}\nonumber \\
 &&+\,
  a_3\,( \SuperField{L}^f \cdot\SuperField{H}_u)\,  \SuperField{\nu}^{\chargec 3} \, 
 \frac{\<\SingletHcB_{f}\> }{\MSingletHc} 
 +\;
\dots\; .
\end{eqnarray}
$\MSingletHi$ correspond to the masses of some Froggatt-Nielsen fields \cite{Froggatt:1980sz} 
which are integrated out
 and produce the effective operators.  
In leading order the neutrino Yukawa matrix  
is thus given by
\vspace{1mm}
\begin{eqnarray}\label{eq:LOYnu} \label{eq:Ynu}
Y^0_\nu
= 
\left(\begin{array}{ccc}
\displaystyle a_1 \frac{\<\SingletHaB_{1}\>}{\MSingletHa}&\displaystyle a_2 \frac{\<\SingletHbB_{1}\>}{\MSingletHb}
&\displaystyle a_3 \frac{\<\SingletHcB_{1}\>}{\MSingletHc}
\\[4mm]
\displaystyle  a_1 \frac{\<\SingletHaB_{2}\>}{\MSingletHa}&\displaystyle  a_2 \frac{\<\SingletHbB_{2}\>}{\MSingletHb}
&\displaystyle  a_3 \frac{\<\SingletHcB_{2}\>}{\MSingletHc}\\[4mm]
\displaystyle a_1 \frac{\<\SingletHaB_{3}\> }{\MSingletHa}&\displaystyle a_2 \frac{\<\SingletHbB_{3}\>}{\MSingletHb}
&\displaystyle a_3 \frac{\<\SingletHcB_{3}\>}{\MSingletHc}
\end{array}
\right)  
.\vspace{1mm}
\end{eqnarray}
The Majorana mass term for the right-handed neutrinos is restricted by the 
discrete symmetries and for our choice of $Z_2$ symmetries   
given by
\begin{eqnarray}\label{eq:MR0withSO(3)}
 \mathcal{W}_{M_\mathrm{RR}} & =& 
\frac{1}{2} \SuperField{\nu}^{\chargec 1} M_{\mathrm{R}1}\SuperField{\nu}^{\chargec 1}
+ 	\frac{1}{2} \SuperField{\nu}^{\chargec 2} M_{\mathrm{R}2}\SuperField{\nu}^{\chargec 2}
+ 	\frac{1}{2} \SuperField{\nu}^{\chargec 3} M_{\mathrm{R}3}\SuperField{\nu}^{\chargec 3}\nonumber \\
&&+ \frac{1}{2} \sum_{i,j,I,J=1}^3\SuperField{\nu}^{\chargec i} (M'_\mathrm{RR})_{ij}\SuperField{\nu}^{\chargec j}\,
\,\delta_{iI}\,\delta_{jJ}\,\sum_{f=1}^3  \frac{\<(\SingletHB_I)_{f}\>
\<(\SingletHB_J)_{f}\>}{M_{N' I} M_{N' J}}
+  \;\dots\; .
\end{eqnarray}
At leading order, the mass matrix is thus forced to a diagonal structure 
\begin{eqnarray}\label{eq:LOMR}
M_\mathrm{RR}^0
=
\begin{pmatrix}
 M_{\mathrm{R}1} &0&0\\
 0&M_{\mathrm{R}2}&0\\
 0&0&M_{\mathrm{R}3}
\end{pmatrix}
.
\end{eqnarray}
In the presence of the $\SO(3)$ symmetry, 
the couplings of  the lepton doublets to the triplet Higgs superfield 
$\SuperField{\Delta}$ are given by
\begin{eqnarray}\label{eq:YDeltaSO(3)}
 \mathcal{W}_{\Delta}  = y_\Delta\, \SuperField{L}^T{}^f \, i\tau_2\,\SuperField{\Delta} \,\SuperField{L}^f 
 +\sum_{f,g=1}^3 \SuperField{L}^T{}^f \,i\tau_2\, \SuperField{\Delta} \,\SuperField{L}^g \, 
 \sum_{I=1}^3 b_I \frac{\<\SingletHiB_{f}\> \<\SingletHiB_{g}\>}{M_{L I}^2}
 + \;\dots\; .
\end{eqnarray}
At leading order, this results in a contribution $Y_\Delta v_\Delta$ 
to the neutrino mass matrix proportional to the unit matrix in flavour space, 
\begin{eqnarray}\label{eq:LOmLL}
Y_{\Delta}^0
=
y_{\Delta} 
\begin{pmatrix}
 1&0&0\\
 0&1&0\\
 0&0&1
\end{pmatrix}
.
\end{eqnarray}
Assuming $M_{NI} = M_{N'I} = M_{LI}$, the next-to-leading order operators 
are of the order 
$(\<\SingletHiB_{f}\> / \MSingletHi)^2 \approx(Y_\nu^0)^2_{If}$. 
If the neutrino Yukawa couplings are very small, the next-to-leading order 
operators are strongly 
suppressed and can approximately be neglected. However, if 
the effective field theory expansion parameters 
$\<\SingletHiB_{f}\> / \MSingletHi$ are relatively large, 
we should stress that a careful analysis of higher-dimensional operators 
of the superpotential and also of the K\"{a}hler potential 
(see e.g.~\cite{King:2003xq})  
has to be performed. Let us consider for example the effect of a 
neutrino Yukawa coupling 
$(Y_\nu^0)_{33} \approx \<\SingletHcB_{3}\> / M_{N 3} 
\approx 0.1$. The next-to-leading 
order operators of equation (\ref{eq:YDeltaSO(3)}) induce a contribution 
to the atmospheric mass squared difference 
$\Delta m^2_\mathrm{atm}:= m_3^2 - m_1^2$ for a given scale of the 
direct mass term 
$m^{\mathrm{II}} \mathbbm{1} = y_\Delta v_\Delta \mathbbm{1}$.  
If we take for example $m^{\mathrm{II}}\approx 0.1$ eV and 
set $b_3= y_\Delta$ and $a_3=1$ for simplicity, this would 
induce a contribution to $\Delta m^2_\mathrm{atm}$ of about 
$2 \cdot 10^{-4}$ eV$^2$, which is still about an order of magnitude below 
the observed experimental value and thus provides only a rather small correction.  
Since we will not assume particularly large neutrino Yukawa couplings in this work, we 
 will ignore the next-to-leading order operators in the following.  

\subsection{The Mass Matrix of the Charged Leptons}\label{sec:Ye}
Before we turn to classes of models which illustrate the 
mechanism, we discuss the generation of the mass matrix for the 
charged leptons.
The most unrestricted case can be achieved by 
introducing three new additional flavour-triplet Higgs superfields 
$\SingletHei_{f}$ and three additional ${Z}'_2$ symmetries, as 
specified in table \ref{tab:repsSO(3)ChargedLeptons}. 
\begin{table}[h]
\begin{center}
\begin{tabular}{l|cccccc}
	$\vphantom{\sqrt{\big|}}$Field  
	& \(\SuperField{e}^{\ChargeC 1}\) & \(\SuperField{e}^{\ChargeC 2}\)& \(\SuperField{e}^{\ChargeC 3}\)
	& \(\SingletHe_{1}\)& \(\SingletHe_{2}\)& \(\SingletHe_{3}\)
			\\
	\hline				
	\(\vphantom{\sqrt{\big|}} \SO(3)\)
	& \(\boldsymbol{1}\) & \(\boldsymbol{1}\) & \(\boldsymbol{1}\) 
	& \(\boldsymbol{3}\) & \(\boldsymbol{3}\) & \(\boldsymbol{3}\)
		\\	
	\(\vphantom{\sqrt{\big|}}{Z}'^1_2 \)
	& \(-\) & \(+\) & \(+\) 
	& \(-\) & \(+\) & \(+\)	
		\\
	\(\vphantom{\sqrt{\big|}}{Z}'^2_2 \)
	& \(+\) & \(-\) & \(+\) 
	& \(+\) & \(-\) & \(+\)	
		\\
	\(\vphantom{\sqrt{\big|}}{Z}'^3_2 \)
	& \(+\) & \(+\) & \(-\) 
	& \(+\) & \(+\) & \(-\)	
\end{tabular}	
\end{center}
\caption{
\label{tab:repsSO(3)ChargedLeptons} 
Representations of the charged leptons and new $G_{321}$-singlets 
under the horizontal symmetries $\SO(3) 
\times \:({Z}'_2)^3$, leading to the most unrestricted scenario.
 The fields are singlets under the symmetries ${Z}^1_2,{Z}^2_2$ and 
 ${Z}^3_2$.  
}
\end{table}	
This results in a general charged lepton Yukawa matrix 
generated by higher-dimensional operators, 
which can be hierarchical from a hierarchy 
of the vevs  $\<\SingletHeaB_{f}\>,\<\SingletHebB_{f}\>$ and 
$\<\SingletHecB_{f}\>$ or from the masses of the Froggatt-Nielsen fields. 
 The superpotential operators for the Yukawa interactions are 
\begin{eqnarray}\label{eq:effectiveYe}
\mathcal{W}_{Y_e} &=& 
- \,a'_1 \, (\SuperField{L}^f\cdot\SuperField{H}_d)\, \SuperField{e}^{\chargec 1} \, 
 \frac{\<\SingletHeaB_{f}\> }{\MSingletHea}
 -
 a'_2\, (\SuperField{L}^f\cdot\SuperField{H}_d) \, \SuperField{e}^{\chargec 2} \, 
 \frac{\<\SingletHebB_{f}\> }{\MSingletHeb}\nonumber \\
 &&-\,
  a'_3\, (\SuperField{L}^f\cdot\SuperField{H}_d)\,  \SuperField{e}^{\chargec 3} \, 
 \frac{\<\SingletHecB_{f}\> }{\MSingletHec} 
 -
\;\dots \; .
\end{eqnarray}
The ${\cal O}(1)$-coefficients $ a_1,a_2,a_3 $ and $ a'_1,a'_2,a'_3 $ 
stem from the realization of the effective operators and are in principle 
calculable within an underlying full theory. 
The leading order Yukawa matrix for
the charged leptons is given by 
\vspace{1mm}
\begin{eqnarray} \label{eq:Ye}
Y^0_e
= 
\left(\begin{array}{ccc}
\displaystyle a'_1 \frac{\<\SingletHeaB_{1}\>}{\MSingletHea} &\displaystyle a'_2 \frac{\<\SingletHebB_{1}\>}{\MSingletHeb}
 &\displaystyle a'_3 \frac{\<\SingletHecB_{1}\>}{\MSingletHec}
 \\[4mm]
\displaystyle  a'_1 \frac{\<\SingletHeaB_{2}\>}{\MSingletHea} &\displaystyle  a'_2 \frac{\<\SingletHebB_{2}\>}{\MSingletHeb}
 &\displaystyle  a'_3 \frac{\<\SingletHecB_{2}\>}{\MSingletHec} \\[4mm]
\displaystyle a'_1 \frac{\<\SingletHeaB_{3}\> }{\MSingletHea} &\displaystyle a'_2 \frac{\<\SingletHebB_{3}\>}{\MSingletHeb}
&\displaystyle a'_3 \frac{\<\SingletHecB_{3}\>}{\MSingletHec} 
\end{array}
\right)
.\vspace{1mm}
\end{eqnarray}
Different choices of discrete symmetries and flavon fields involved in the 
generation of the charged lepton mass matrix can give  
more predictive scenarios. 
Obviously, the $\SO(3)$ symmetry might leave the quark sector 
completely unaffected. On the other hand, the framework discussed 
so far can also be extended to the quark sector, yielding rather unrestricted 
quark masses with a natural hierarchy among the columns of the quark Yukawa
matrices. For the present however, and in the rest of the paper, 
we restrict ourselves to the lepton sector.

\section{SO(3) Vacuum Alignment}\label{sec:VacuumAlignment}
To stay as minimal as possible, we consider in the following 
the case that the charged leptons 
couple to the the same flavon fields  
 $\SingletH_I$ as the neutrinos. This corresponds to  
replacing ${Z'}_2^{I} \rightarrow Z_2^{I}$ and $\SingletHei_f \rightarrow
\SingletHi_f$ in section \ref{sec:Ye}. 
The Yukawa matrix of the charged leptons is then related to the neutrino Yukawa
matrix. 
The $\SO(3)$ flavour symmetry in the lepton sector is  
spontaneously broken by the vevs 
$\<\SingletHaB_f\>$. 
Let us consider first the general case with 
complex vevs. 
A simultaneous SO(3) rotation of the three vevs allows to eliminate
three real degrees of freedom. 
We will shortly consider a scenario where the SO(3) vacuum is 
aligned such that the
vevs are real. Complex phases in the Yukawa matrices $Y_\nu$
and $Y_e$ then stem entirely from 
the coefficients $ a_1,a_2,a_3 $ and $ a'_1,a'_2,a'_3 $. 
It turns out that with this vacuum alignment, which allows for 
three texture zeros in the Yukawa matrices, the type II scenario 
can realize the observed masses and mixings in the lepton sector in a
particularly natural way.

\subsection{Real Alignment for the SO(3) Vacuum}\label{sec:RealVacuumAlignment}
A possibility to achieve real vevs is to 
introduce three driving superfields   
$\SuperField{A},\SuperField{B}$ and $\SuperField{C} $ and to  
assume a superpotential of the form 
\begin{eqnarray}
\mathcal{W} = 
\SuperField{A} (\SingletH_1^2 - \Lambda_1^2) 
+ \SuperField{B} (\SingletH_2^2 - \Lambda_2^2)
+ \SuperField{C} (\SingletH_3^2 - \Lambda_3^2) 
\end{eqnarray}
with positive soft mass squareds 
\begin{eqnarray}
  m^2_{\theta_1} >0 
 \; , \quad  m^2_{\theta_2}  >0  
\quad \mbox{and} \quad  m^2_{\theta_3}  >0 
\end{eqnarray}
for the scalar components 
$\theta_1,\theta_2$ and $\theta_3$  
of the flavon superfields $\SingletH_1,\SingletH_2$ and $\SingletH_3$ 
 (see e.g.~\cite{Barbieri:1999km}). 
 $\Lambda_1,\Lambda_2$ and $\Lambda_3$ could stem from vevs of some 
 $\SO(3)$-singlets. They  
 can be chosen positive and real by a proper phase choice for these fields.  
Let us explicitly consider the vacuum alignment for  
$\theta_1$. Minimization of $|F_{A}|^2$ yields 
\begin{eqnarray}
\Lambda_1^2=\sum_f (\mathrm{Re}\<\SingletHaB_{f}\>)^2 - 
\sum_i (\mathrm{Im}\<\SingletHaB_{f}\>)^2 + 
2 i\sum_f \mathrm{Re}\<\SingletHaB_{f}\>\mathrm{Im}\<\SingletHaB_{f}\>\: .
\end{eqnarray}
With our choice 
 $\Lambda_1 \in \mathbbm{R}^+$, this leads to the two conditions 
\begin{subequations}\begin{eqnarray}
0 &=& \sum_f \mathrm{Re}\<\SingletHaB_{f}\>\mathrm{Im}\<\SingletHaB_{f}\> \; ,\\
\label{eqn:MinCondFA_1} \Lambda_1^2&=&\sum_f (\mathrm{Re}\<\SingletHaB_{f}\>)^2 - 
\sum_f (\mathrm{Im}\<\SingletHaB_{f}\>)^2\; .
 \end{eqnarray}\end{subequations}
The soft mass term $\mathcal{V}_{s} = m^2_{\theta_1} \theta_1^* \theta_1$ 
 deforms the driving potential and its minimization under the conditions 
 from $|F_{A}|^2$ results in  
 $\mathrm{Im}\<\SingletHaB_{f}\>=0$ for all components $f \in \{1,2,3\}$. 
 This can be seen by plugging  
 equation (\ref{eqn:MinCondFA_1}) into $\mathcal{V}_{s}$, which
 yields  
\begin{eqnarray}
\mathcal{V}_{s} &=& 
 m^2_{\theta_1}\,\left[
\sum_f (\mathrm{Re}\<\SingletHaB_{f}\>)^2 + 
\sum_f (\mathrm{Im}\<\SingletHaB_{f}\>)^2\right] \nonumber \\
&=& 
 m^2_{\theta_1}\,\left[
 \Lambda_1^2 + 
2 \sum_f (\mathrm{Im}\<\SingletHaB_{f}\>)^2 \right] .
\end{eqnarray}
 Thus, the small positive mass squared forces the 
 vev $\<\theta_1\>$ to be real. Treating $\<\theta_2\>$ and $\<\theta_3\>$ 
 analogously, we obtain a real alignment for the vevs of all 
 scalar components of the flavon superfields,
 \begin{eqnarray}
 \mathrm{Im}\<\SingletHB_{1}\>=0 \; , \quad
 \mathrm{Im}\<\SingletHB_{2}\>=0 \; , \quad
 \mathrm{Im}\<\SingletHB_{3}\>=0 \; . 
 \end{eqnarray} 
 The moduli of the vev vectors are
 given by 
 \begin{eqnarray}
 |\<\SingletHB_{1}\>|^2 \approx \Lambda_1^2 \; , \quad
 |\<\SingletHB_{2}\>|^2 \approx \Lambda_2^2 \; , \quad
 |\<\SingletHB_{3}\>|^2 \approx \Lambda_3^2 \; , 
 \end{eqnarray}
 which could lead to a hierarchy among the vevs via a hierarchy among  
 $\Lambda_1$, $\Lambda_2$ and $\Lambda_3$.  

We can now use the $\SO(3)$ freedom to set two components of one of the 
vev vectors, say $\<\SingletHiB_{f}\>$, and one corresponding component of a
second vev vector, say $\<\SingletHjB_{f}\>$, to zero. 
Defining the real expansion parameters 
\begin{subequations}\label{eq:DefEpsilon}\begin{eqnarray}
&&\varepsilon_I:=|a_I|\frac{\Lambda_I}{\MSingletHi}\; , \;\;
\varepsilon_J:=|a_J|\frac{\Lambda_J}{\MSingletHj}\; ,
\;\;\varepsilon_K:=|a_K|\frac{\Lambda_K}{\MSingletHk}\; , \\
&&\varepsilon'_I:=|a'_I|\frac{\Lambda_I}{\MSingletHei}\; , \;\;
\varepsilon'_J:=|a'_J|\frac{\Lambda_J}{\MSingletHej}\; ,\;\;
\varepsilon'_K:=|a'_K|\frac{\Lambda_K}{\MSingletHek}\; , \;\;
\end{eqnarray}\end{subequations}
without loss of
generality we can write 
\begin{subequations}\label{eq:VacAlignmentColumns}\begin{eqnarray}
a_I \frac{\<\SingletHB_I\>}{\MSingletHi} \!\!\!&:=&\!\!\! \left(\!\begin{array}{c} 0 \\  0 \\ h \,e^{i \delta_I} \varepsilon_I\end{array}\!\!\right) \!, \;
a_J\frac{\<\SingletHB_J\>}{\MSingletHj} :=           \!     \left(\!\begin{array}{c} 0 \\  e\,e^{i \delta_J}\varepsilon_J \\ f \,e^{i\delta_J}\varepsilon_J \end{array}\!\!\right)\! , \;
a_K \frac{\<\SingletHB_K\>}{\MSingletHk} :=  \!              \left(\!\begin{array}{c} a \,e^{i \delta_K}\varepsilon_K \\  b\,e^{i\delta_K}\varepsilon_K \\ c \,e^{i \delta_K}\varepsilon_K
\end{array}\!\!\right)\! ,  \\
a'_I \frac{\<\SingletHB_I\>}{\MSingletHei} \!\!\!&:=&\!\!\! \left(\!\begin{array}{c} 0 \\  0 \\ h \,e^{i \delta'_I}\varepsilon'_I \end{array}\!\!\right) \!,  \;
a'_J \frac{\<\SingletHB_J\>}{\MSingletHej} :=                \left(\!\begin{array}{c} 0 \\  e\,e^{i \delta'_J}\varepsilon'_J \\ f \,e^{i\delta'_J}\varepsilon'_J \end{array}\!\!\right) \!, \;
a'_K \frac{\<\SingletHB_K\>}{\MSingletHek} :=  \!             \left(\!\begin{array}{c} a \,e^{i \delta'_K}\varepsilon'_K \\  b\,e^{i\delta'_K}\varepsilon'_K \\ c\,e^{i \delta'_K}\varepsilon'_K
\end{array}\!\!\right)\!,
\end{eqnarray}\end{subequations}
with real coefficients $\{a,b,c,e,f,h\}$ satisfying 
\begin{eqnarray}
&&h \approx 1 \; , \;\; e^2 + f^2 \approx 1 \; , \;\; a^2 + b^2 + c^2 \approx 1 \; .
\end{eqnarray}
We choose the convention that the labels $\{1,2,3\}$ are assigned to 
$I,J$ and $K$ such that 
\begin{eqnarray}
\varepsilon'_3 > \varepsilon'_2 > \varepsilon'_1 
\end{eqnarray} 
holds in the charged lepton sector. 
This also defines the labels of $\varepsilon_1,\varepsilon_2$ and 
$\varepsilon_3$ and of the heavy 
mass eigenvalues $M_{\mathrm{R}1},M_{\mathrm{R}2}$ 
and $M_{\mathrm{R}3}$ of the right-handed neutrinos.  
 In the following, by a global phase transformation of all the leptons, we
arrange for the direct mass term for the
neutrinos proportional to the unit matrix to be real and positive. In addition,
we absorb possible phases of $M_{\mathrm{R}1},M_{\mathrm{R}2}$ 
and $M_{\mathrm{R}3}$ in the columns of $Y_\nu$.
The phases in the Yukawa matrices then stem from the coefficients $a_1,a_2,a_3$ 
and $a'_1,a'_2,a'_3$ which are in general complex, 
\begin{subequations}\begin{eqnarray}
&&\da := \mbox{Arg} \,(a_1) \; ,\; \; \db := \mbox{Arg} \,(a_2) \; ,\; \;\dc :=
\mbox{Arg} \,(a_3) \; ,\; \; \\
&&\daP := \mbox{Arg} \,(a'_1) \; ,\; \; \dbP := \mbox{Arg} \,(a'_2) \; ,\; \;\dcP
:= \mbox{Arg} \,(a'_3) \; .
\end{eqnarray}\end{subequations}

\subsection{Textures for the Yukawa Matrices and Type II Scenarios} 

The parameterization of the the SO(3)-breaking vacuum specified in equation
(\ref{eq:VacAlignmentColumns}) uses the basis where 
the Yukawa matrices $Y_\nu$ and $Y_e$ each have three zero entries. 
Motivation for this choice of basis would come from  a full theory beyond 
the framework presented here. 
Table \ref{tab:TypeIIScenarios} shows the 
textures for $Y_\nu$ and $Y_e$ for different proportions 
of $\varepsilon'_I, \varepsilon'_J$ and $\varepsilon'_K$ using the ordering
convention $\varepsilon'_3 > \varepsilon'_2 > \varepsilon'_1$. 
Additional characteristic features of the Yukawa matrices are that 
 each column has a common complex phase and that, 
 without additional symmetries, the non-zero components of  
 each column have a common typical order of magnitude defined by 
 the expansion parameters $\varepsilon_1,\varepsilon_2,\varepsilon_3$ and 
 $\varepsilon'_1,\varepsilon'_2,\varepsilon'_3$.   
 The hierarchy among the masses of the 
charged leptons can naturally be realized via  
$\varepsilon'_3\gg\varepsilon'_2\gg\varepsilon'_1$, which we will assume 
 in the following. 

\begin{table}[p]
\begin{eqnarray*}
\begin{array}{|c|c|c|}
\hline
\mbox{Model}&Y^0_\nu&Y^0_e\\
\hline
\mbox{A1} 
&
\left(\begin{array}{ccc}
a\,e^{i \da}\epsa&0&0\\
b\,e^{i \da}\epsa&e\,e^{i \db}\epsb&0\\
c\,e^{i \da}\epsa&f\,e^{i \db}\epsb&h\,e^{i \dc}\epsc
\end{array}
\right)
&
\left(\begin{array}{ccc}
a\,e^{i \daP}\epsaP&0&0\\
b\,e^{i \daP}\epsaP&e\,e^{i \dbP}\epsbP&0\\
c\,e^{i \daP}\epsaP&f\,e^{i \dbP}\epsbP&h\,e^{i \dcP}\epscP
\end{array}
\right)\\
\hline
\mbox{A2} 
&
\left(\begin{array}{ccc}
0&a\,e^{i \db}\epsb&0\\
e\,e^{i \da}\epsa&b\,e^{i \db}\epsb&0\\
f\,e^{i \da}\epsa&c\,e^{i \db}\epsb&h\,e^{i \dc}\epsc
\end{array}
\right)
&
\left(\begin{array}{ccc}
0&a\,e^{i \dbP}\epsbP&0\\
e\,e^{i \daP}\epsaP&b\,e^{i \dbP}\epsbP&0\\
f\,e^{i \daP}\epsaP&c\,e^{i \dbP}\epsbP&h\,e^{i \dcP}\epscP
\end{array}
\right)\\
\hline
\mbox{B1} 
&
\left(\begin{array}{ccc}
a\,e^{i \da}\epsa&0&0\\
b\,e^{i \da}\epsa&0&e\,e^{i \dc}\epsc\\
c\,e^{i \da}\epsa&h\,e^{i \db}\epsb&f\,e^{i \dc}\epsc
\end{array}
\right)
&
\left(\begin{array}{ccc}
a\,e^{i \daP}\epsaP&0&0\\
b\,e^{i \daP}\epsaP&0&e\,e^{i \dcP}\epscP\\
c\,e^{i \daP}\epsaP&h\,e^{i \dbP}\epsbP&f\,e^{i \dcP}\epscP
\end{array}
\right)\\
\hline 
\mbox{B2} 
&
\left(\begin{array}{ccc}
0&a\,e^{i \db}\epsb&0\\
0&b\,e^{i \db}\epsb&e\,e^{i \dc}\epsc\\
h\,e^{i \da}\epsa&c\,e^{i \db}\epsb&f\,e^{i \dc}\epsc
\end{array}
\right)
&
\left(\begin{array}{ccc}
0&a\,e^{i \dbP}\epsbP&0\\
0&b\,e^{i \dbP}\epsbP&e\,e^{i \dcP}\epscP\\
h\,e^{i \daP}\epsaP&c\,e^{i \dbP}\epsbP&f\,e^{i \dcP}\epscP
\end{array}
\right)\\
\hline
\mbox{C1} 
&
\left(\begin{array}{ccc}
0&0&a\,e^{i \delta_3}\,\varepsilon_3   \\
0&e\,e^{i \delta_2}\,\varepsilon_2&b\,e^{i \delta_3}\,\varepsilon_3 \\
h\,e^{i \delta_1}\,\varepsilon_1&f\,e^{i \delta_2}\,\varepsilon_2&c\,e^{i \delta_3}\,\varepsilon_3
\end{array}
\right)
&
\left(\begin{array}{ccc}
0&0&a\,e^{i \delta'_3}\,\varepsilon'_3   \\
0&e\,e^{i \delta'_2}\,\varepsilon'_2&b\,e^{i \delta'_3}\,\varepsilon'_3 \\
h\,e^{i \delta'_1}\,\varepsilon'_1&f\,e^{i \delta'_2}\,\varepsilon'_2&c\,e^{i\delta'_3}\,\varepsilon'_3
\end{array}
\right)
\\
\hline
\mbox{C2} 
&
\left(\begin{array}{ccc}
0&0&a\,e^{i \delta_3}\,\varepsilon_3   \\
e\,e^{i \delta_1}\,\varepsilon_1&0&b\,e^{i \delta_3}\,\varepsilon_3 \\
f\,e^{i \delta_1}\,\varepsilon_1&h\,e^{i \delta_2}\,\varepsilon_2&c\,e^{i \delta_3}\,\varepsilon_3
\end{array}
\right)
&
\left(\begin{array}{ccc}
0&0&a\,e^{i \delta'_3}\,\varepsilon'_3   \\
e\,e^{i \delta'_1}\,\varepsilon'_1&0&b\,e^{i \delta'_3}\,\varepsilon'_3 \\
f\,e^{i \delta'_1}\,\varepsilon'_1&h\,e^{i \delta'_2}\,\varepsilon'_2&c\,e^{i\delta'_3}\,\varepsilon'_3
\end{array}
\right)
\\
\hline
\end{array}
\end{eqnarray*}
\caption{
\label{tab:TypeIIScenarios} 
Textures for the leading order Yukawa matrices from 
real vacuum alignment with three zero entries. The 
motivation for choosing such a basis would come a full theory 
 beyond the scope of the framework presented here. 
We use the convention that the right-handed lepton fields multiply the Yukawa
matrices from the right, as defined in equation (\ref{eq:effectiveYnu}) 
and (\ref{eq:effectiveYe}).  
Our sorting of the vev vectors and labeling of the 
columns of $Y^0_\nu$ and $Y^0_e$ is such that 
$\varepsilon'_3>\varepsilon'_2>\varepsilon'_1$. 
The mass matrix $M^0_\mathrm{RR}$ of the right-handed neutrinos is diagonal 
 (see equation (\ref{eq:MR0withSO(3)})) and the direct mass term for 
 the neutrinos is given by $m^\mathrm{II}\mathbbm{1}$ at leading order. 
This leads to type II see-saw models with either small mixing 
from the charged leptons and bi-large mixing from the neutrinos (A1 and A2),  
large atmospheric mixing from the charged
leptons (B1 and B2) or models where in principle all mixing can be produced via the
charged lepton mass matrix (C1 and C2). 
The neutrino mass squared differences and the contributions to the lepton 
mixings from the type I neutrino mass 
matrix depend crucially on the ratios 
${\epsa^2}/{M_{\mathrm{R}1}},{\epsb^2}/{M_{\mathrm{R}2}}, 
{\epsc^2}/{M_{\mathrm{R}3}}$. The hierarchy among the masses of the 
charged leptons can naturally be realized via a hierarchy 
$\varepsilon'_3\gg\varepsilon'_2\gg\varepsilon'_1$. 
}
\end{table}

\subsubsection{Models Type A: Large Mixing $\theta_{23}$ from the Neutrino Mass
Matrix}
In the models of type A, the mixing $\theta^e_{23}$ 
from the charged lepton mass matrix $M_e$ is approximately zero. 
The almost maximal total lepton mixing $\theta_{23}$ thus has to be 
generated in the neutrino mass matrix. 
This can for example be achieved if the dominant contribution to the 
type I part of the neutrino mass matrix stems from the right-handed 
neutrino $\nu_{\mathrm{R}}^{ 2}$ for model A1 or $\nu_{\mathrm{R}}^{ 1}$ 
for model A2. 
As in the single right handed neutrino dominance case 
 \cite{King:1998jw,King:1999cm}   
for a pure type I neutrino mass matrix, the  
 condition for a nearly maximal $\theta_{23}$ is 
 $|e| \approx |f|$. 
In addition, the zero in the first element of the column containing the
coefficients $e$ and $f$ in general avoids the generation of a large 
$\theta^\nu_{13}$ from the neutrino mass matrix.  
Model A1 has the additional feature that 
$\theta^e_{12}$ and $\theta^e_{13}$ are
very small, whereas in model type A2 a charged lepton mixing 
$\tan (\theta^e_{12}) = a/b$ is generated which induces a contribution to 
the total lepton mixings $\theta_{12}$ and $\theta_{13}$.  
In order not to violate the CHOOZ bound on $\theta_{13}$, 
$\theta^e_{12}$ has to be rather small which means that $a$ has to be
somewhat smaller than $b$.  
Consequently, the large mixing $\theta_{12} \approx 32^\circ$ should be
generated mainly in the neutrino sector.  
With sequential right-handed
neutrino dominance \cite{King:1999mb,King:2002nf} for
$m^\mathrm{I}_{\mathrm{LL}}$, this can easily be
realized. We will study model A1 in detail in section \ref{sec:Typw2SeqRhdNuDom}.

\subsubsection{Models Type B: Large Mixing $\theta_{23}$ from the Charged Leptons}
The mixing $\theta^e_{23}$ from the charge lepton mass matrix is given by 
$\tan (\theta^e_{23}) \approx e/f$. $e \approx f$ can explain the 
nearly maximal total lepton mixing $\theta_{23}$ in a lopsided mass model,
provided that the contribution $\theta^\nu_{23}$ is small, which can e.g.~be
achieved if $\nu_{\mathrm{R}}^{ 2}$ is the 
dominant right-handed neutrino 
for model B1 and $\nu_{\mathrm{R}}^{ 1}$ 
for model B2. Since as for the model A2, a large $\theta^e_{12}$ in model 
B2 would 
induce a large contribution to $\theta_{13}$, the large solar mixing should be
generated by the neutrino mass matrix.

\subsubsection{Models Type C: Large Angles for all the Charged Lepton Mixings}
For models of type C, it is possible to generate bi-large 
lepton mixing entirely from $Y_e$. 
Small mixing from the neutrino mass matrix would correspond to  
 $\nu_{\mathrm{R}}^{ 1}$ being the dominant right-handed neutrino
and $\nu_{\mathrm{R}}^{ 2}$ being subdominant for model C1 or to 
$\nu_{\mathrm{R}}^{ 1}$ being dominant 
and $\nu_{\mathrm{R}}^{ 2}$ being subdominant for model C2. 
For the model C2, the large mixings are 
 given by $\tan (\theta_{12})=a/b$ and $\tan
 (\theta_{23})=\sqrt{a^2+b^2}/c$ (see also \cite{Babu:2001cv}). 
While $\theta_{13}\approx 0$ for model C2, 
it has to be taken care that $\theta_{13}$ 
is below the experimental bounds in model C1.

\section{A Type II Scenario with Sequential RHND
}\label{sec:Typw2SeqRhdNuDom}
As an example for a type II model where the bi-large lepton mixing stems 
from the neutrino mass matrix, we now consider explicitly the model A1 
of table \ref{tab:TypeIIScenarios}. We make the additional  
assumption of sequential righthanded neutrino dominance
(RHND) \cite{King:1999mb,King:2002nf}, leading to a hierarchical type 
I part $m_\nu^{\mathrm{I}}$ of the neutrino mass matrix. 
The vev structure for this case allows to define 
(see equation (\ref{eq:VacAlignmentColumns}))
 \begin{subequations}\begin{eqnarray}
a_3 \frac{\<\SingletHB_3\>}{\MSingletHc} \!\!\!&:=&\!\!\! \!\left(\begin{array}{c} 0 \\  0 \\ h \,e^{i \delta_3} \varepsilon_3\end{array}\right) \!, \;\; \!
a_2\frac{\<\SingletHB_2\>}{\MSingletHb} :=  \!              \left(\begin{array}{c} 0 \\  e\,e^{i \delta_2}\varepsilon_2 \\ f \,e^{i\delta_2}\varepsilon_2 \end{array}\right)\! , \;\;\! 
a_1 \frac{\<\SingletHB_1\>}{\MSingletHa} := \!               \left(\begin{array}{c} a \,e^{i \delta_1}\varepsilon_1 \\  b\,e^{i\delta_1}\varepsilon_1 \\ c \,e^{i \delta_1}\varepsilon_1
\end{array}\right)\! ,  \\
a'_3 \frac{\<\SingletHB_3\>}{\MSingletHec} \!\!\!&:=&\!\!\! \!\left(\begin{array}{c} 0 \\  0 \\ h \,e^{i \delta'_3}\varepsilon'_3 \end{array}\right) \!,  \;\;\!
a'_2 \frac{\<\SingletHB_2\>}{\MSingletHeb} :=                \left(\begin{array}{c} 0 \\  e\,e^{i \delta'_2}\varepsilon'_2 \\ f \,e^{i\delta'_2}\varepsilon'_2 \end{array}\right) \!, \;\;\! 
a'_1 \frac{\<\SingletHB_1\>}{\MSingletHea} := \!              \left(\begin{array}{c} a \,e^{i \delta'_1}\varepsilon'_1 \\  b\,e^{i\delta'_1}\varepsilon'_1 \\ c\,e^{i \delta'_1}\varepsilon'_1
\end{array}\right)\!,
\end{eqnarray}\end{subequations}
leading to the Yukawa matrices 
\begin{eqnarray}\label{eqn:YukawaMatrices_A1}
Y^0_\nu
= 
\left(\begin{array}{ccc}
a\,e^{i \da}\epsa&0&0\\
b\,e^{i \da}\epsa&e\,e^{i \db}\epsb&0\\
c\,e^{i \da}\epsa&f\,e^{i \db}\epsb&h\,e^{i \dc}\epsc
\end{array}
\right)
   \!, \;\;
 Y^0_e
= 
\left(\begin{array}{ccc}
a\,e^{i \daP}\epsaP&0&0\\
b\,e^{i \daP}\epsaP&e\,e^{i \dbP}\epsbP&0\\
c\,e^{i \daP}\epsaP&f\,e^{i \dbP}\epsbP&h\,e^{i \dcP}\epscP
\end{array}
\right)
 \!.
\end{eqnarray}
In order to match notation with reference \cite{King:1999mb,King:2002nf}, we define 
\begin{eqnarray}
M^0_\mathrm{RR}
=:
\left(\begin{array}{ccc}
\!X&0&0\!\\
\!0&Y&0\!\\
\!0&0&X'\!
\end{array}
\right)\!,
\end{eqnarray}
denoting the mass of the dominant right-handed neutrino by $Y$ and the mass of
the subdominant one by $X$.  
The sequential RHND condition we impose is then 
\begin{eqnarray}\label{eq:SequSubDominace}
\left|\frac{\epsb^2}{Y}\right| \gg 
\left|\frac{\epsa^2}{X}\right|\gg 
\left|\frac{\epsc^2}{X'}\right| \; .
\end{eqnarray}
The leading order type II neutrino mass matrix is then given by 
\begin{eqnarray}\label{eq:TypeIIMassMatrix_A1}
m_{\mathrm{LL}}^\nu 
 &\!\!\!=\!\!\!& 
 m^\mathrm{II}
\!\left(\begin{array}{ccc}
\!1&\!0\!&0\!\\
\!0&\!1\!&0\!\\
\!0&\!0\!&1\!
\end{array}
\right)\!
 -
\frac{e^{i2\db} \epsb^2 v_u^2}{Y}  \! \left(\begin{array}{ccc}
\!0&0\!&0\!\\
\!0&e^2\!&ef\!\\
\!0&ef\!&f^2\!
\end{array}
\right)\!
-
\frac{e^{i2\da} \epsa^2 v_u^2}{X}\! \left(\begin{array}{ccc}
\vphantom{f^2}\!a^2&ab&ac\!\vphantom{f^2}\\
\vphantom{f^2}\!ab&b^2&bc\!\vphantom{f^2}\\
\vphantom{f^2}\!ac&bc&c^2\!\vphantom{f^2}
\end{array}
\right)\! . 
\end{eqnarray}
We now analyze the lepton masses, mixings and CP phases which are generated 
by $m^\nu_{\mathrm{LL}}$ and by the lepton mass matrix $M_e=v_d Y_e$
analytically.

\subsection{Analytic Results for Neutrino Masses, Lepton Mixings and CP Phases}
The mixing matrix in the lepton sector, the MNS matrix $U_{\mathrm{MNS}}$, 
is defined by the charged electroweak current 
$\overline{{e}_\mathrm{L}}^f \gamma^\mu  U_{MNS} {\nu}^f_\mathrm{L}$ 
in the mass basis.  
Defining the diagonalization matrices $U_{e_\mathrm{L}},U_{e_\mathrm{R}}$ and 
$U_{\nu_\mathrm{L}}$ by  
\begin{eqnarray}
U_{e_\mathrm{L}} \, M_e \,U^\dagger_{e_\mathrm{R}} =
\left(\begin{array}{ccc}
\!m_e&0&0\!\\
\!0&m_\mu&0\!\\
\!0&0&m_\tau\!
\end{array}
\right)\! , \quad
U_{\nu_\mathrm{L}} \,m^\nu_{\mathrm{LL}}\,U^T_{\nu_\mathrm{L}} =
\left(\begin{array}{ccc}
\!m_1&0&0\!\\
\!0&m_2&0\!\\
\!0&0&m_3\!
\end{array}
\right)\! ,
\end{eqnarray}
the MNS matrix is given by
\begin{eqnarray}
U_{\mathrm{MNS}} = U_{e_\mathrm{L}} U^\dagger_{\nu_\mathrm{L}}\; .
\end{eqnarray}
We use the parameterization 
$
U_{\mathrm{MNS}} = R_{23} U_{13} R_{12} P_0
$ 
with $R_{23}, U_{13}, R_{12}$ and $P_0$ being defined as 
\begin{align}
R_{12}:=
\left(\begin{array}{ccc}
  c_{12} & s_{12} & 0\\
  -s_{12}&c_{12} & 0\\
  0&0&1\end{array}\right)
  , \:&
\quad U_{13}:=\left(\begin{array}{ccc}
   c_{13} & 0 & s_{13}e^{-i\delta}\\
  0&1 & 0\\
  - s_{13}e^{i\delta}&0&c_{13}\end{array}\right)  ,  \nonumber \\ 
R_{23}:=\left(\begin{array}{ccc}
 1 & 0 & 0\\
0&c_{23} & s_{23}\\
0&-s_{23}&c_{23}
 \end{array}\right)
  , \:&
 \quad P_0:= 
 \begin{pmatrix}
 1&0&0\\0&e^{i\beta_2}&0\\0&0&e^{i\beta_3}
  \end{pmatrix}. 
\end{align}

\noindent The masses of the charged leptons are given by 
 $m_\tau = h \,\epscP\, v_d,m_\mu = e \,\epsbP\, v_d$  and 
 $m_e = a \,\epsaP\, v_d$.
Unless $h,e$ or $a$ are zero,  
we can easily choose $\epscP,\epsbP$ and $\epsaP$ such that the right charged
lepton masses are produced. 
In addition we note that 
 the mixings $\theta^e_{12},\theta^e_{13}$ and $\theta^e_{23}$, which stem from 
 $U_{e_\mathrm{L}}$ and could
 contribute to the MNS matrix, are very small. 
 Furthermore, in leading order each column of $M_e$ has a common complex phase, which can be 
 absorbed by $U_{e_\mathrm{R}}$. Therefore, the
 charged leptons do not influence the leptonic CP phases in this approximation.

\begin{figure}[t]
   \begin{center}
 $\vcenter{\hbox{\includegraphics[scale=1]{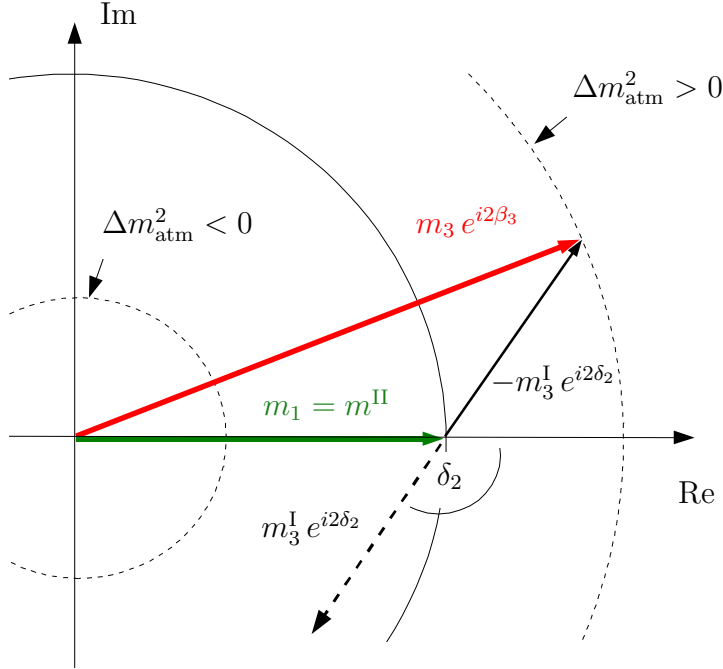}}}$
 \end{center}
 \caption{\label{fig:MassesinComplexPlane}
Graphical illustration of how $m_3^{\mathrm{I}}$ has to be chosen in oder to
produce the experimentally observed atmospheric mass squared difference 
$\Delta m^2_\mathrm{atm}$. In the complex plane, 
$m_3 e^{i 2 \beta_3}= m_1 - m_3^{\mathrm{I}} 
e^{i 2 \delta_2}$ has to lie on one of the dashed circles which correspond to values
of $m_3$ such that the right $|\Delta m^2_\mathrm{atm}|:= |m_3^2 - m_1^2|$ 
 is produced.   
The outer dashed circle corresponds to 
$\Delta m^2_\mathrm{atm}>0$ and a so-called normal mass ordering $m_3 > m_1$, 
while the inner 
 dashed circle corresponds to 
$\Delta m^2_\mathrm{atm}<0$ and an inverse ordering $m_3 < m_1$. 
An analogous picture can be
drawn for  $m_2 e^{i 2 \beta_2}= m_1 - m_2^{\mathrm{I}} e^{i 2 \delta_1}$. Note 
that for $\Delta m^2_\mathrm{sol}$ and $\theta_{12}$, only the combinations   
$\Delta m^2_\mathrm{sol}>0$ with $\theta_{12}< 45^\circ$ and 
$\Delta m^2_\mathrm{sol}<0$ with $\theta_{12}> 45^\circ$ are allowed 
by experiment. From the low energy point of view 
the two possibilities are equivalent, however they correspond to different model parameters at
high energy.}
\end{figure}

Using the analytical methods for diagonalizing neutrino mass matrices with small $\theta_{13}$
derived in \cite{King:2002nf}, from 
$m^\nu_{\mathrm{LL}} = m_{\mathrm{LL}}^{\mathrm{II}} + 
m_{\mathrm{LL}}^{\mathrm{I}}$ we find for the mixing 
angles
 \begin{subequations}\begin{eqnarray}
\label{AnalyticResultForT23} \tan (\theta_{23}) &\approx& \frac{|e|}{|f|} \; , \\
\label{AnalyticResultForT13} \tan (2 \theta_{13}) &\approx& 
\frac{2 \,|a|\,\epsa^2\,v_u^2}{X} \,
\frac{|\sin (\theta_{23}) |b| + \cos (\theta_{23}) |c| \,\mbox{sign}\,(b\, c\,
e\, f)|}{
|2\,  m^{\mathrm{II}}\, \sin (\widetilde \delta) + m_3^{\mathrm{I}}
e^{i(2\db + 3\pi/2  -\widetilde \delta)}|
}  \; ,\\
\label{AnalyticResultForT12}\tan (\theta_{12}) &\approx& \frac{|a|}{|\cos (\theta_{23}) |b| - 
\sin (\theta_{23}) |c|\,\mbox{sign}\,(b\, c\,
e\, f)| } \; ,
\end{eqnarray}\end{subequations} 
where $m_i^{\mathrm{I}}$ ($i\in \{1,2,3\}$) are the mass eigenvalues of the hierarchical
$m_{\mathrm{LL}}^{\mathrm{I}}$ given by
 \begin{subequations}\begin{eqnarray}
m_1^{\mathrm{I}} &=&  {\cal O} \left(\frac{\epsc^2 v_u^2}{X'}\right)   \;\approx\;0 \; , \vphantom{\frac{|e|}{|f|}}\\
\label{eq:m2I_A1} m_2^{\mathrm{I}} &\approx& \frac{(|a|^2 + |\cos (\theta_{23}) |b| - 
\sin (\theta_{23}) |c|\,\mbox{sign}\,(b\, c\,
e\, f)|^2) \,\epsa^2\,v_u^2}{X}
\;\approx\;\frac{|a|^2\,\epsa^2\,v_u^2 }{\sin^2 (\theta_{12}) X}\;,\\
m_3^{\mathrm{I}} &\approx& \frac{(|e| \sin (\theta_{23})+|f|\cos (\theta_{23}))^2\,\epsb^2\,v_u^2 }{Y}\;, 
\end{eqnarray}\end{subequations} 
and with $\widetilde \delta$ defined by 
\begin{eqnarray}\label{eq:DiracCP_A1_1}
 \tan (\widetilde \delta) := 
\frac{m_3^{\mathrm{I}} \, \sin (2 \db -  2 \da)}{  
m_3^{\mathrm{I}} \,\cos (2 \db -  2 \da) - 2 \, m^{\mathrm{II}} \, \cos (2 \da)}\, .
\end{eqnarray}
Given $\tan (\widetilde \delta)$, $\widetilde \delta$ has to be chosen such that 
\begin{eqnarray}
\frac{\sin (\theta_{23}) |b| + \cos (\theta_{23}) |c| \,\mbox{sign}\,(b\, c\,
e\, f)}{
\mbox{sign}\,(a\, b) \, [
2\,  m^{\mathrm{II}}\,e^{-i (2\da+ 3\pi/2)} \, \sin (\widetilde \delta) + m_3^{\mathrm{I}}
e^{-i(2\da -2\db+\widetilde \delta)}
]
}
\ge 0 \;.
\end{eqnarray}
This does not effect $\theta_{13}$, which we have defined to be $\ge 0$, however
it is relevant for extracting the Dirac CP phase $\delta$, 
given by 
\begin{eqnarray}\label{eq:DiracCP_A1_2}
\delta &\approx& 
\left\{
\begin{array}{cl}
\widetilde \delta & \mbox{for $P\ge0$\,,}\\
\widetilde \delta + \pi & \mbox{for $P<0$\,,}
\end{array}
\right.
\end{eqnarray} 
with $P$ being defined by
\begin{eqnarray}
P := \frac{
\cos (\theta_{23}) |b| - 
\sin (\theta_{23}) |c|\,\mbox{sign}\,(b\, c\,
e\, f)
}{
\mbox{sign}\,(a\, b) \, [
(\cos (\theta_{23}) |b| - 
\sin (\theta_{23}) |c|\,\mbox{sign}\,(b\, c\,
e\, f))^2 - |a|^2
]
}\; .
\end{eqnarray}
The mass eigenvalues of the complete type II neutrino mass matrix are given by
 \begin{subequations}\label{eq:ComplMassEigenvOfMnuTypeII}\begin{eqnarray}
m_1  &\approx&  |m^{\mathrm{II}}|\;,\\
m_2  &\approx& | m^{\mathrm{II}} - m_2^{\mathrm{I}} \,e^{i 2 \da}|\;,\\
m_3  &\approx& | m^{\mathrm{II}} - m_3^{\mathrm{I}} \,e^{i 2 \db}|\;, 
\end{eqnarray} \end{subequations} 
and, for $m^{\mathrm{II}} \not= 0$, the Majorana phases $\beta_2$ and $\beta_3$  
can be extracted by
\begin{subequations}\label{eq:MajPhases_A1}\begin{eqnarray}
\beta_2  &\approx& \frac{1}{2}\mbox{arg}\,( m^{\mathrm{II}} - m_2^{\mathrm{I}} \,e^{i 2 \da})\;,\\
\beta_3  &\approx& \frac{1}{2}\mbox{arg}\,( m^{\mathrm{II}} - m_3^{\mathrm{I}} \,e^{i 2 \db})\;.
\end{eqnarray}\end{subequations} 
A graphical illustration of the correlation between the masses and
phases is given in figure \ref{fig:MassesinComplexPlane}. 

\subsection{Discussion}
\begin{figure}[t]
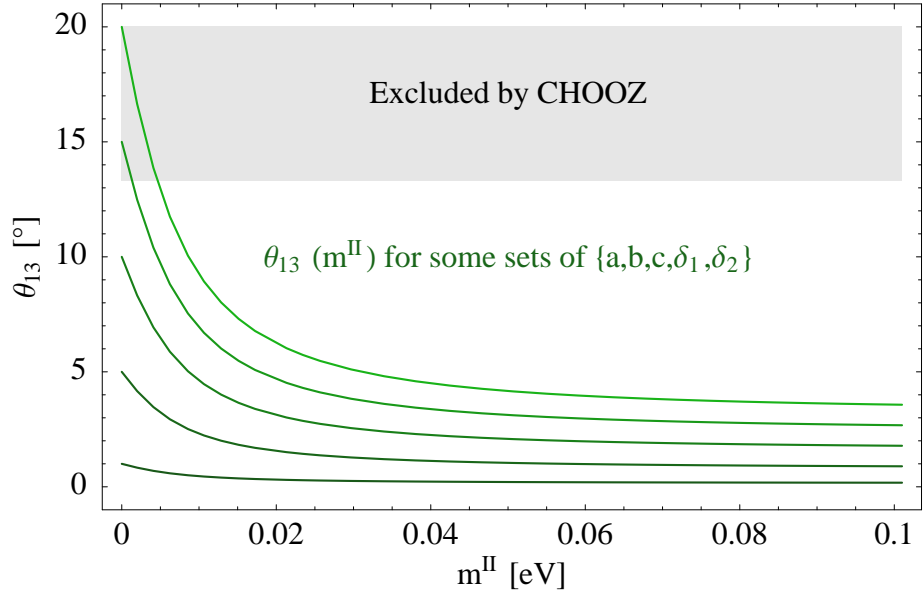

\begin{center}
  {$\CenterEps[1]{Theta13vsmII_2}$}
 \caption{\label{fig:Theta13vsmII}
Dependence of $\theta_{13}$ on the type II mass scale $m^\mathrm{II}$ for some 
sets of parameters $\{a,b,c,\da,\db\}$. Adding
the unit matrix contribution $m^{\mathrm{II}}_{\mathrm{LL}}=m^\mathrm{II}
\mathbbm{1}$ leads to a decrease of $\theta_{13}$ as $m^\mathrm{II}$ 
increases. 
The grey range for $\theta_{13}$ is excluded by experiment at $3 \sigma$.  
 }
\end{center}
\end{figure}

\begin{figure}
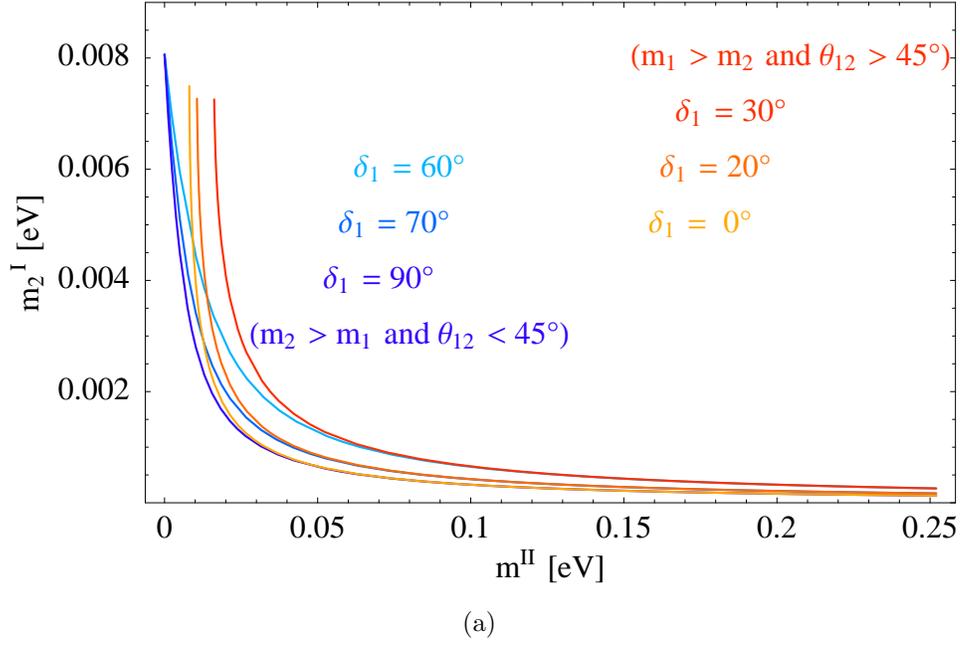
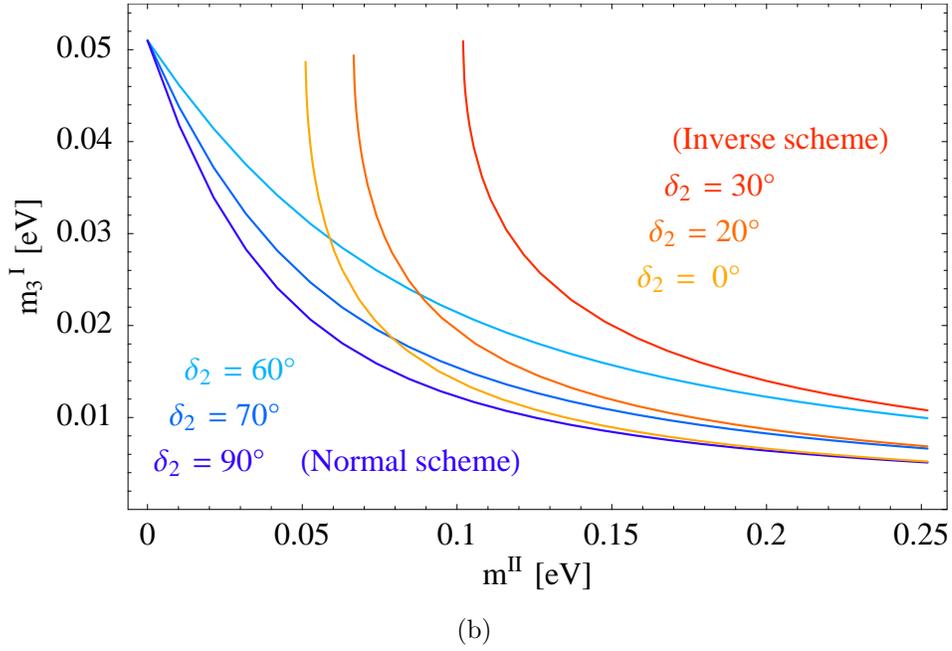

\begin{center}
  \subfigure[]{$\CenterEps[1]{m2I_A1}$} \\
  \subfigure[]{$\CenterEps[1]{m3I_A1}$} 
 \caption{\label{fig:m2Iandm3I}
Values for $m_2^{\mathrm{I}}$ and $m_3^{\mathrm{I}}$ as functions of
$m^\mathrm{II}$, required in order to
produce the best-fit values $\Delta m^2_\mathrm{atm}\approx 2.6 \cdot
10^{-3}$ and $\Delta m^2_\mathrm{sol}\approx 6.9 \cdot 10^{-5}$ for some 
typical choices of the complex phases $\da$ and $\db$. For
a graphical illustration how $m_2^{\mathrm{I}}$ and $m_3^{\mathrm{I}}$ are
determined, see figure \ref{fig:MassesinComplexPlane}. 
 } 
\end{center}
\end{figure}

\begin{figure}
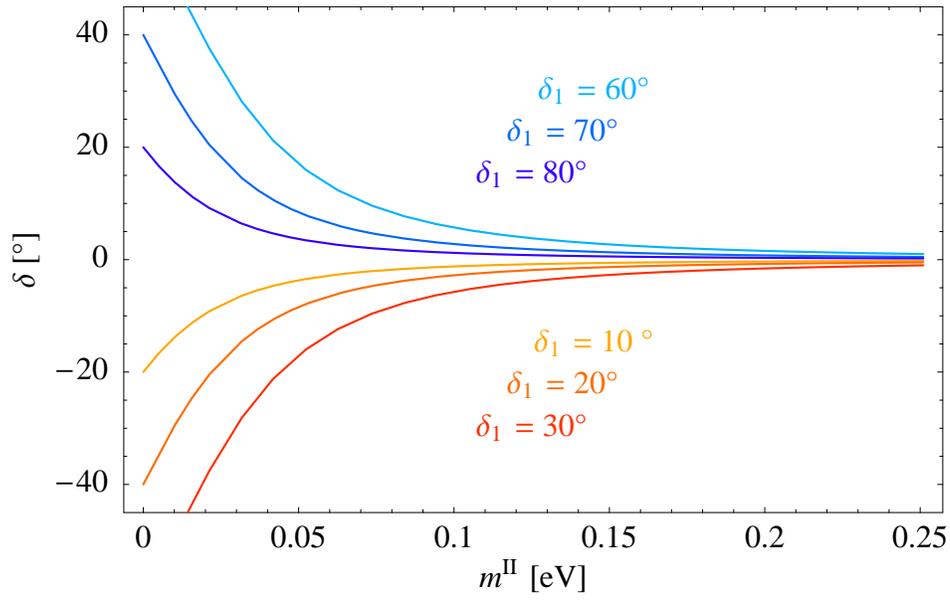

 \begin{center}
 \CenterEps[1]{DiracCPphase_A1}
 \end{center}
 \caption{\label{fig:DiracCPphase_A1}
Dependence of the Dirac CP phase $\delta$ on $m^\mathrm{II}$  for some typical
values for the complex phase $\da$ and with 
$\delta_2 = 90^\circ$. A large type II mass scale $m^\mathrm{II}$ generically 
implies a small Dirac CP phase.  
 }
\end{figure}

We find that for the mixings, only $\theta_{13}$ is affected by the 
direct mass term proportional to the unit matrix, whereas $\theta_{12}$ and 
$\theta_{23}$ are independent of $m^\mathrm{II}$. 
This is due to the fact that $\theta_{23}$ only depends on the dominant contribution 
to the type I part of the neutrino mass matrix and 
$\theta_{12}$ depends only on the subdominant type I part to leading order in $\theta_{13}$.  
On the other hand, $\theta_{13}$ depends on both, the dominant and the 
subdominant part.  
Looking at equation (\ref{AnalyticResultForT13}) for $\theta_{13}$, 
it naively seems as 
$\theta_{13}$ goes to zero as $m^\mathrm{II}$ increases. However, by using the
result for $\widetilde \delta$, the denominator 
of equation (\ref{AnalyticResultForT13}) can be rewritten as
\begin{eqnarray}
|2\,  m^{\mathrm{II}}\, \sin (\widetilde \delta) + m_3^{\mathrm{I}}
e^{i(2\db + 3\pi/2  -\widetilde \delta)}| = \left| m_3^{\mathrm{I}}\,
\frac{\cos (2 \db - \widetilde \delta)}{\cos (2 \da)}\right| ,
\end{eqnarray}
which does no longer depend on $m^\mathrm{II}$ explicitly. 
Note that the term $\cos (2 \db - \widetilde \delta)/\cos (2 \da)$ goes to 
$1$ for small $m^\mathrm{II}$ and to $\cos (2 \db)/\cos (2 \da)$ for large 
$m^\mathrm{II}$, because $ \widetilde \delta$ goes to $0$ as discussed later and 
is shown in figure \ref{fig:DiracCPphase_A1}.\footnote{
Note that there are special choices of the complex phases where 
 mass eigenvalues of the type I part of the neutrino mass matrix do not become smaller   
  for larger $m^\mathrm{II}$. 
This is e.g.~the case for $\db=45^\circ$, as can be seen from figure 
\ref{fig:MassesinComplexPlane}. In this case $\widetilde \delta$ does 
not go to zero as $m^\mathrm{II}$ increases. 
For large $m^\mathrm{II}$, we regard this special choices 
for the phases as unnatural, since two large quantities from $m^{\mathrm{II}}_{\mathrm{LL}}$ 
and $m^{\mathrm{II}}_{\mathrm{LL}}$ have to 
conspire in order to produce a small mass squared difference. 
} 
Besides the additional explicit dependence on the parameters $a,b$ and $c$,  
the mixing angle $\theta_{13}$ is only suppressed by a factor of 
$m_2^\mathrm{I}/m_3^\mathrm{I}$. 
However, as we discuss later and 
is shown in figure \ref{fig:m2Iandm3I}, $m_2^\mathrm{I}$ gets smaller for
increasing $m^\mathrm{II}$ much faster that $m_3^\mathrm{I}$. Consequently, 
$\theta_{13}$ in fact decreases with increasing $m^\mathrm{II}$, however it does
not go to zero for large $m^\mathrm{II}$. The dependence of $\theta_{13}$ on the
type II mass scale is
illustrated in figure \ref{fig:Theta13vsmII}.

As the type II contribution 
$m^{\mathrm{II}}_\mathrm{LL}$ gets larger and the neutrino mass scale increases, 
the mass splittings have to get smaller (see e.g.~figure \ref{fig:NuMasses}) 
in order to match 
the experimentally observed mass squared 
differences $\Delta m^2_\mathrm{sol}:= m_2^2 - m_1^2$ and 
$|\Delta m^2_\mathrm{atm}|:= |m_3^2 - m_1^2|$. 
Depending on $m^\mathrm{II}$ and the
complex phases $\da$ and $\db$, this determines the eigenvalues 
$m_2^\mathrm{I}$ and $m_3^\mathrm{I}$ of the type I part of the neutrino 
mass matrix (see figure \ref{fig:MassesinComplexPlane}). 
Note that with a hierarchical type I part $m^{\mathrm{I}}_{\mathrm{LL}}$, 
a normal mass
ordering as well as an inverse ordering can be achieved. The latter is 
only possible if $(m^\mathrm{II} \cos (2 \db))^2 > |\Delta m^2_\mathrm{atm}|$. 
The dependence of $m_2^\mathrm{I}$ and $m_3^\mathrm{I}$ on the type II mass scale 
$m^{\mathrm{II}}$ is shown in figure \ref{fig:m2Iandm3I}. 
For producing the small mass squared differences for a larger
$m^\mathrm{II}$ in a natural way, i.e.~without cancellations of two relatively 
large terms $m^{\mathrm{I}}_{\mathrm{LL}}$ and $m^{\mathrm{II}}_{\mathrm{LL}}$, $m_2^\mathrm{I}$ 
and $m_3^\mathrm{I}$ have to be smaller. 
From equation (\ref{eq:ComplMassEigenvOfMnuTypeII}) and the definitions of 
$m_2^{\mathrm{I}}$ and $m_3^{\mathrm{I}}$ 
we find that the masses $Y$ and $X$ of the right-handed neutrinos, 
which control these mass 
splittings, can be chosen appropriately in a natural way.
As we have already noted, the ratio $m_2^\mathrm{I}/m_3^\mathrm{I}$ for a 
partially degenerate mass spectrum is smaller than for hierarchical neutrino masses. 
This means that in the type II scenario with partially degenerate neutrino masses, 
sequential righthanded neutrino dominance is even more natural than 
in the pure type I see-saw case.

\begin{figure}
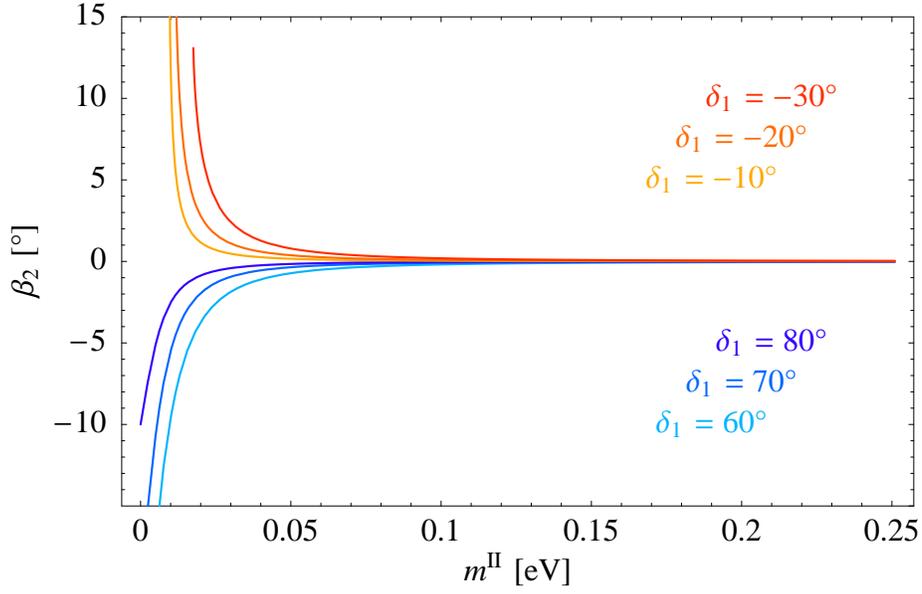
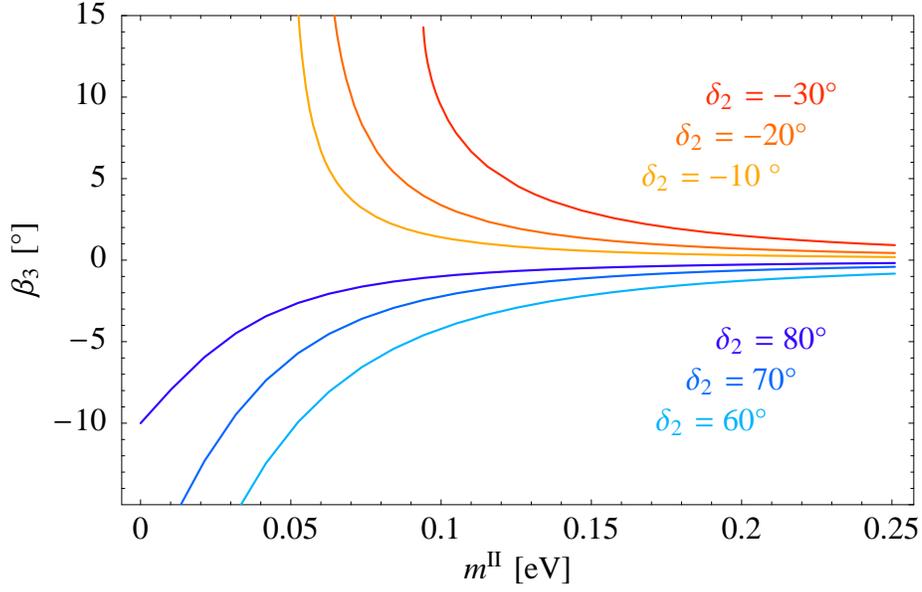

\begin{center}
  \subfigure[]{$\CenterEps[1]{Beta2_A1}$} \\
  \subfigure[]{$\CenterEps[1]{Beta3_A1}$} 
 \caption{\label{fig:MajPhases}
Diagram (a) shows the dependence of the Majorana CP phase $\beta_2$ on 
$m^\mathrm{II}$  for some typical
values for the complex phase $\da$ and with $\delta_2 = 90^\circ$. Diagram 
(b) shows the Majorana CP phase $\beta_3$ as a function of 
$m^\mathrm{II}$ for some sample values $\db$ and with $\delta_1 = 90^\circ$. 
We see that both phases get smaller for larger $m^\mathrm{II}$. The CP phase 
$\beta_2$ is
below $5^\circ$ already for $m^\mathrm{II}$ larger than about $0.03$ eV. 
 } 
\end{center}
\end{figure}

As can be seen from equation (\ref{eq:MajPhases_A1}), the $m^{\mathrm{II}}$-dependence
of $m_2^{\mathrm{I}}$ and $m_3^{\mathrm{I}}$ implies that the 
Majorana phases $\beta_1$ and $\beta_2$ get smaller for larger 
$m^{\mathrm{II}}$ (see figure \ref{fig:MajPhases}). 
From equations (\ref{eq:DiracCP_A1_1}) and (\ref{eq:DiracCP_A1_2}), 
we conclude that the Dirac CP phase $\delta$ generically gets smaller for 
larger $m^{\mathrm{II}}$ as well. 
Quantitatively, the dependence of $\delta$ on 
the type II mass scale is shown in figure \ref{fig:DiracCPphase_A1} for some
sample choices of $\da$ and $\db$.

\clearpage\newpage

\section{Renormalization Group Corrections}\label{sec:RGRunning}
In order to compare the predictions of see-saw scenarios with the 
experimental data obtained at low energy, the 
renormalization group (RG) running of the effective neutrino
mass matrix has to be taken into account.  
It is known that for partially degenerate neutrino masses, 
RG corrections to the neutrino mixing angles  
 can be significant.   
The corrections to the masses and mass squared 
differences are relevant even for strongly hierarchical neutrino masses 
and they can be enhanced or suppressed for a partially degenerate neutrino 
mass spectrum depending on the CP phases. 

Let us assume for example that the considered type II scenarios are 
embedded into a unified model 
where all additional degrees of freedom except for the ones contained in the 
right-handed neutrino superfields $\SuperField{\nu}_{\mathrm{R}}^{ \Nf}$ are
integrated out above some energy scale $M_\mathrm{U}$. At $M_\mathrm{U}$, which
could be the scale of gauge coupling unification, the model parameters are defined.  
In this case, the 
effective neutrino mass matrix has to be run from $M_\mathrm{U}$ to low 
energy using the $\beta$-functions for the various energy 
 ranges above and between the see-saw scales \cite{Antusch:2002rr,Antusch:2002ek}
 and below the mass scale of the lightest right-handed neutrino   
\cite{Chankowski:1993tx,Babu:1993qv,Antusch:2001ck,Antusch:2001vn,Antusch:2002ek}.

For an accurate computation of the RG corrections for the neutrino mass parameters,
the coupled system of differential equations has to be solved successively 
for the various effective theories \cite{King:2000hk,Antusch:2002rr}. 
In addition to the parameters of the MNS matrix, the RG running between
$M_\mathrm{U}$ and the electroweak scale $M_{\mathrm{EW}}$ then depends on 
the additional degrees of freedom corresponding to the lepton Yukawa couplings 
and the masses of the right-handed neutrinos.   

For a small difference of the Majorana CP phases corresponding to the mass
eigenvalues $m_1$ and $m_2$, 
which is generically the case in the presented type II framework, 
radiative corrections to the lepton 
mixings have a characteristic property, as has been pointed out in 
\cite{Antusch:2002hy,Antusch:2002fr}. The running of
the solar mixing $\theta_{12}$ is generically larger than the RG corrections to 
the other mixings, if the atmospheric mixing is large already at high energy. 
We will see that in particular $\theta_{12}$ can be significantly 
lower at high energy, depending on $m^{\mathrm{II}}$ and $\tan \beta$.

We will now estimate the generic size of the RG corrections for the 
mixing angles by making
additional simplifications and assumptions. At first, let us consider the case
that the heaviest right-handed neutrino corresponds to the column of the neutrino
Yukawa matrix with the largest entries and that it has a mass larger than
$M_\mathrm{U}\approx 2\cdot 10^{16}$. 
If we then assume that the other entries of
the neutrino Yukawa matrix are very small compared to 
$y_\tau$, we can approximately neglect the running due to the neutrino Yukawa 
matrix and, for a rough estimate, simply use the RGEs for the neutrino mass
operator 
below the see-saw scales.   
In order to see which quantities control the size of the RG effects, it is 
useful to consider the RGEs for the parameters of the MNS matrix
\cite{Chankowski:1999xc,Casas:1999tg,Antusch:2003kp}. 
For example, the running of
the mixing angles in leading order in the small quantity $\theta_{13}$ in the MSSM
is given by \cite{Antusch:2003kp}   
\vspace{1mm}
\begin{eqnarray} \label{eq:AnalyticApproxT12}
\frac{\mathrm{d}}{\mathrm{d} t}\, \theta_{12}
& \!=\! &
        -\frac{y_\tau^2}{32\pi^2} \,
        \sin 2\theta_{12} \, s_{23}^2\, 
        \frac{
      | {m_1} + {m_2}\, e^{\I  2 \beta_2}|^2
     }{\Delta m^2_\mathrm{sol} }     
        + \mathscr{O}(\theta_{13}) \;,
         \label{eq:Theta12Dot}
\end{eqnarray} 
\begin{eqnarray} \label{eq:AnalyticApproxT13} 
\frac{\mathrm{d}}{\mathrm{d} t}\,\theta_{13}
& \!=\! & 
        \frac{y_\tau^2}{32\pi^2} \, 
        \sin 2\theta_{12} \, \sin 2\theta_{23} \,
        \frac{m_3}{\Delta m^2_\mathrm{atm} \left( 1+\zeta \right)}
        \times
\nonumber\\
&& \quad \times
        \left[
         m_1 \cos(2\beta_3-\delta) -
         \left( 1+\zeta \right) m_2 \, \cos(2\beta_3-2\beta_2-\delta) -
         \zeta m_3 \, \cos\delta
        \right] \nonumber \\
	&&
        +       \mathscr{O}(\theta_{13}) \;,
 \label{eq:Theta13Dot}
\end{eqnarray}
\begin{eqnarray} \label{eq:AnalyticApproxT23} 
\frac{\mathrm{d}}{\mathrm{d} t}\, \theta_{23}
& \!=\! &
        -\frac{y_\tau^2}{32\pi^2} \, \sin 2\theta_{23} \,
        \frac{1}{\Delta m^2_\mathrm{atm}} 
        \left[
         c_{12}^2 \, |m_2\, e^{\I 2 (\beta_3-\beta_2)} + m_3|^2 +
         s_{12}^2 \, \frac{|m_1\, e^{\I 2 \beta_3} + m_3|^2}{1+\zeta}
        \right] 
        \nonumber\\
        & & {}
        + \mathscr{O}(\theta_{13}) \;,
 \label{eq:Theta23Dot}\vspace{1mm}
\end{eqnarray} 
where, with the renormalization scale $\mu$, $t$ is defined by 
$t:=\ln (\mu/\mu_0)$  
and where we have used the abbreviation 
$\zeta := {\Delta m^2_\mathrm{sol}}/{\Delta 
m^2_\mathrm{atm}}$. Compared to the SM, $y_\tau$ in the MSSM 
is given by $y_\tau=y_\tau^\mathrm{SM} \sqrt{ 1+\tan^2\beta }$, 
which yields an enhancement of the RG effects for large $\tan \beta$. 
In addition, the running is generically enhanced due to factors of the type 
$m^2_i/\Delta m^2_\mathrm{sol}$ or $m^2_i/\Delta m^2_\mathrm{atm}$   
if the neutrino mass scale is larger than the
mass differences. For partially degenerate neutrinos, this is in particular 
the case for $m^2_i/\Delta m^2_\mathrm{sol}$, which enhances the running of 
$\theta_{12}$.

\begin{figure}[p]
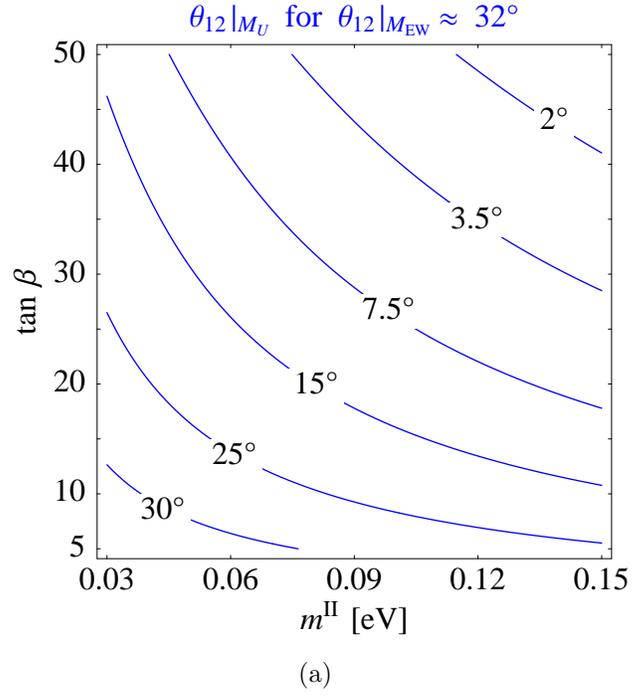
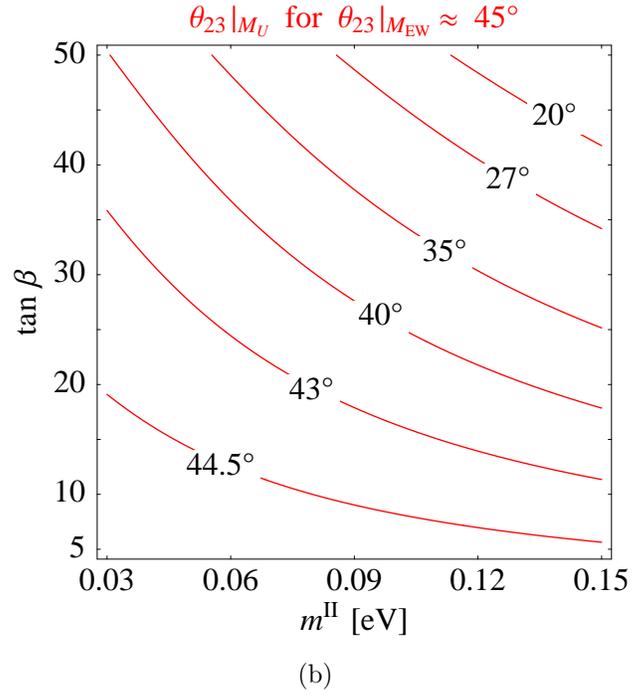

\begin{center}
  \subfigure[]{$\CenterEps[1]{RunningTheta12_2}$}\\
  \subfigure[]{$\CenterEps[1]{RunningTheta23_2}$} 
 \caption{\label{fig:RGCorrectionsMixingAngles}
RG corrected values  
$\theta_{12}|_{M_\mathrm{U}}$ and $\theta_{23}|_{M_\mathrm{U}}$ at high energy, 
required in order to produce the best-fit values  
$\theta_{12} \approx 32^\circ$ and 
$\theta_{23} \approx 45^\circ$ at low energy $M_{\mathrm{EW}}$. We have
considered the case of a normal mass ordering, i.e.~$\Delta m^2_\mathrm{atm}>0$.
 The used approximations and assumptions are explained in the text.   
 }
\end{center}
\end{figure}

\begin{figure}[p]
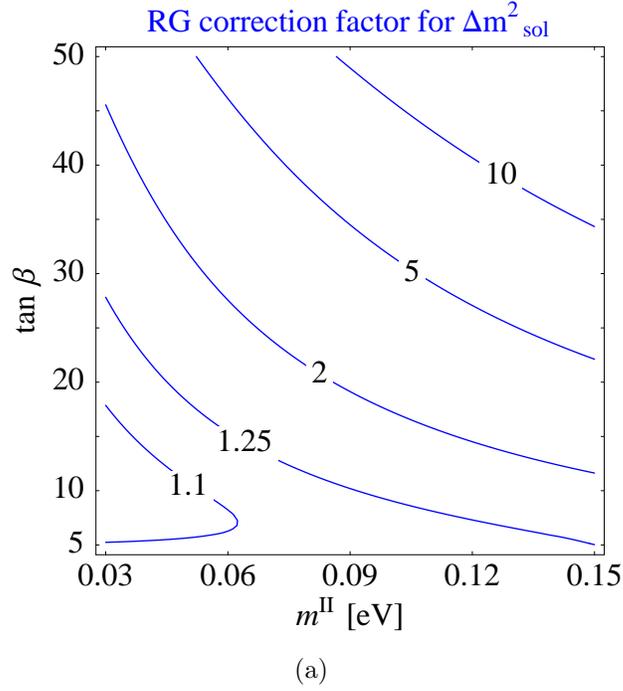
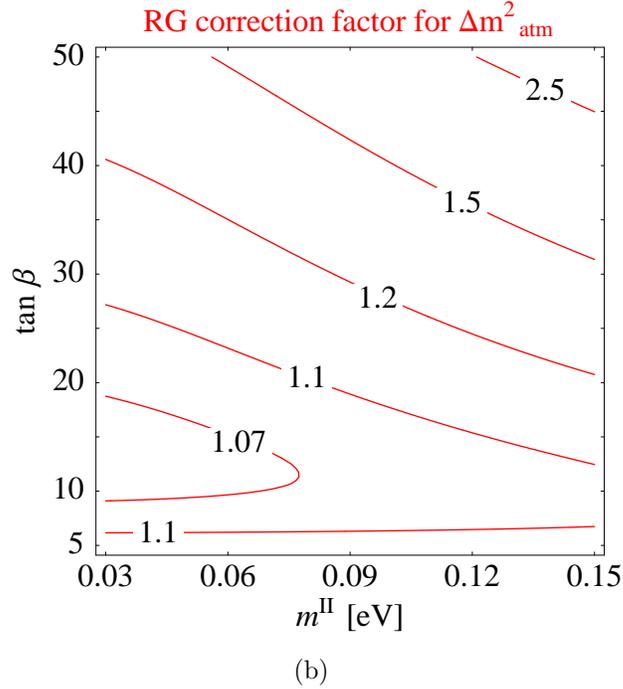

\begin{center}
  \subfigure[]{$\CenterEps[1]{Runningm2Dsol_2}$}\\
  \subfigure[]{$\CenterEps[1]{Runningm2Datm_2}$} 
 \caption{\label{fig:RGCorrectionsMassSquaredDiffs}
RG correction factors for  
$\Delta m^2_\mathrm{sol}|_{M_\mathrm{U}}$ and 
$\Delta m^2_\mathrm{atm}|_{M_\mathrm{U}}$ with respect 
to the low energy best-fit values   
%
for the case of a normal mass ordering, i.e.~$\Delta m^2_\mathrm{atm}>0$.
For the parameter region shown in the plots, both mass squared differences are
larger at high energy.  
The used approximations and assumptions are explained in the text.   
 }
\end{center}
\end{figure}

For an estimate of the RG corrections, we have solved the 
evolution of the parameters numerically from low energy to high energy 
under the additional
assumptions that $\theta_{13}|_{M_\mathrm{EW}} = 0^\circ$ 
and $\beta_3|_{M_\mathrm{EW}} = \beta_2|_{M_\mathrm{EW}} = 0^\circ$. 
Below the SUSY breaking scale, which we have taken to be $1$ TeV, we assume the
theory to be effectively the Standard Model. 
The values for
$\theta_{12}|_{M_\mathrm{U}}$ and $\theta_{23}|_{M_\mathrm{U}}$ required 
in order to produce the best fit value
$\theta_{12}|_{M_\mathrm{EW}}\approx 32^\circ$ and 
$\theta_{23}|_{M_\mathrm{EW}}\approx 45^\circ$ are shown in figure 
\ref{fig:RGCorrectionsMixingAngles}. 
The RG correction factors for  
$\Delta m^2_\mathrm{sol}$ and $\Delta m^2_\mathrm{atm}$ with respect 
to the low energy values  
$\Delta m^2_\mathrm{sol} \approx 6.9 \cdot 10^{-5}$ eV$^2$ and 
$\Delta m^2_\mathrm{atm} \approx 2.6 \cdot 10^{-3}$ eV$^2$
are shown in figure \ref{fig:RGCorrectionsMassSquaredDiffs} for the case of a
normal mass ordering, i.e.~$\Delta m^2_\mathrm{atm}>0$.

It is known that instability under radiative corrections can require  
unnatural fine-tuning of the high energy parameters in order to 
produce the experimentally observed mass squared differences and mixings at low energy. 
Indeed, this is the case for nearly degenerate neutrino masses close to the
experimental upper bounds and 
in addition large $\tan \beta$, such that the slopes  
$\frac{\mathrm{d}}{\mathrm{d} t}\theta_{12}$ and 
$\frac{\mathrm{d}}{\mathrm{d} t}\Delta m^2_\mathrm{sol}$ become very large.  
As we can see from figures 
\ref{fig:RGCorrectionsMixingAngles} and 
\ref{fig:RGCorrectionsMassSquaredDiffs}, for the considered ranges of $\tan \beta \in [5,50]$ and 
$m^\mathrm{II}\in [0.03,1.5]$ eV, 
 the RG corrections are 
well behaved and do not require any fine-tuning of the high energy parameters.  
Depending on $\tan\beta$, the RG effects for a
given neutrino mass scale can either lead to 
significantly changed values for the mixings and mass squared difference 
at high energy or cause only rather small corrections. We conclude that though
the requirement of naturalness restricts the neutrino mass scale for a given 
$\tan \beta$ via stability arguments in our scenarios, this does not lead to any
problems for partially degenerate neutrinos. Nevertheless, the running has to be
included in a careful analysis and it can easily be estimated using the 
figures \ref{fig:RGCorrectionsMixingAngles} and 
\ref{fig:RGCorrectionsMassSquaredDiffs}.

\clearpage

\section{Neutrinoless Double Beta Decay}

We now discuss the implications of the type II see-saw scenarios 
with spontaneously broken SO(3) flavour symmetry for $00\nu\beta$ decay. 
In order to be explicit, we focus on the models 
of type A1 and B1 with sequential right-handed neutrino dominance 
and vacuum alignment as introduced in section 
\ref{sec:Typw2SeqRhdNuDom}.  
 The effective mass $\Braket{m_\nu}$ for $00\nu\beta$ decay is 
 then 
 given by
\begin{eqnarray}\label{eq:0nbbDecayMass_A1}
\Braket{m_\nu} \;=\;
\left|\, \sum_i (U_\mathrm{MNS})_{1i}^2 \, m_i \,\right|
\;=\; \left| (m^\nu_\mathrm{LL})_{11} \right|
\;\approx\; \left|\,m^{\mathrm{II}} -
\sin^2 (\theta_{12}) \,  m_2^{\mathrm{I}} \, e^{2 i \da} 
\,\right|   ,
\end{eqnarray} 
as can be seen from equations (\ref{eq:TypeIIMassMatrix_A1}) and 
(\ref{eq:m2I_A1}). 
A discussion of the RG corrections to 
$\Braket{m_\nu}$ can be found in \cite{Antusch:2003kp}.
The dependence of $\Braket{m_\nu}$ on the direct mass term 
$m_\nu^{\mathrm{II}}=m^{\mathrm{II}}\mathbbm{1}$ and on 
the complex phase $\da$ is illustrated in figure 
\ref{fig:EffMass0nbbDecay}. 
  For the models A1 and B1, we obtain 
$< m_\nu > \approx m_\nu^{\mathrm{II}}$ already for a lightest neutrino mass
of about $0.02$ eV. For all the models with a real vacuum alignment, the general 
statement holds that when the direct mass term $m^{\mathrm{II}}\mathbbm{1}$ 
becomes dominant over the type I part of the mass matrix, 
the effective mass  
$< m_\nu >$ is approximately given by $m^{\mathrm{II}}$.  
The cancellations which 
 can occur in the presence of a large difference of the Majorana phases
 associated with $m_1$ and $m_2$ are then absent and thus there are good 
 prospects are for detecting $\Braket{m_\nu}$ in these classes of type II 
 see-saw models.

\begin{figure}
\begin{center}
  \subfigure[]{$\CenterEps[1]{EffMass0nbbDecay2}$} \\
  \subfigure[]{$\CenterEps[1]{EffMass0nbbDecay}$} 
 \caption{\label{fig:EffMass0nbbDecay}
The diagrams (a) and (b) show the effective mass  $< m_\nu >$ for neutrinoless 
double beta decay and the ratio 
$< m_\nu > / m^{\mathrm{II}}$  
as functions of the type II mass scale $m^{\mathrm{II}}$ for some typical
choices of the complex phase $\da$ in models of type A1 and B1 
(see table \ref{tab:TypeIIScenarios}).  
For other values of $\da$, the curves lie within the grey region of diagram (b).
 We see that $< m_\nu > \approx m^{\mathrm{II}}$ already for $m^\mathrm{II}$
 larger than about $0.02$ eV.  
We have required that the best-fit value for the 
solar mass squared
difference $\Delta m^2_\mathrm{sol}$ is produced in a natural way, 
i.e.~without cancellation of two large 
terms in equation (\ref{eq:0nbbDecayMass_A1}), and we have used the best-fit
experimental value for $\theta_{12}$.   
 }
\end{center}
\end{figure}

\section{Discussion and Conclusions}
We have proposed a type II upgrade of type I see-saw models leading 
to new classes 
of models where partially degenerate neutrinos are as natural as 
hierarchical ones. 
A spontaneously broken SO(3) flavour symmetry forces the direct mass term
for the neutrinos to be proportional to the unit matrix at leading order. 
This allows in principle to boost the 
 mass of the lightest neutrino 
 to any desired value leading from
 hierarchical to nearly degenerate mass spectra. 
Naturalness of models with nearly degenerate neutrino masses consists of two 
issues: First there is tree-level naturalness, which means that all parameters 
and in particular the small mass squared differences and bi-large mixing are 
produced by the model in a natural way. We have shown that our framework 
is natural in this respect and could produce any desired 
level of degeneracy at tree-level.  
Second, there is naturalness with respect to RG corrections. The latter   
should not require fine-tuning of the high-energy model
parameters in order to produce the low-energy experimental values. RG
corrections are un-suppressed in our scenario and thus, depending on $\tan
\beta$, the naturalness 
requirement restricts the neutrino mass scale. 
Instead of nearly degenerate neutrinos, we have therefore considered 
partially degenerate neutrinos with a mass scale up to 
about $0.15$ eV, where the running does not lead to 
any naturalness problems. This mass range is particularly interesting since it 
might be accessible to experiments on neutrinoless double beta decay.

For breaking the SO(3) flavour symmetry, we have considered a minimal 
set of flavon fields and a real alignment mechanism for their 
SO(3)-breaking vevs. The real vacuum 
alignment implies that there exists a basis where the Yukawa matrices have three
texture zeros.
This leads to classes of type II see-saw models with either small mixing from the charged
leptons (type A), almost maximal atmospheric mixing from the charged
leptons (type B) or models where in principle all mixing can be produced via the
charged lepton mass matrix (type C). Characteristic features of them are  
that each column of the Yukawa matrices has a common complex phase and that, 
 without additional symmetries, the non-zero components of  
 each column have a common typical order of magnitude. In addition, 
 for partially degenerate neutrino masses where the type II
 contribution $m^{\mathrm{II}}$ dominates over the type I part, the Majorana phases 
 associated with the neutrino mass eigenvalues are predicted to be 
 small and the 
 effective mass for neutrinoless double beta decay is approximately given 
 by $m^{\mathrm{II}}$. Future experiments on neutrinoless double beta decay can thus 
in principle rule out or confirm partially degenerate neutrino masses 
in this framework.  

Bi-large neutrino mixing and the two small mass squared differences can 
 naturally be realized with sequential right-handed 
neutrino dominance \cite{King:1999mb,King:2002nf} for the type I 
contribution to the neutrino mass matrix. One of the right-handed neutrinos and the
corresponding column of the neutrino Yukawa matrix gives the dominant
contribution to the type I mass matrix and is responsible for the large 
atmospheric mixing $\theta_{23}$ and the mass squared difference 
$\Delta m^2_\mathrm{atm}:= m_3^2 - m_1^2 $. The subdominant right-handed neutrino and the
corresponding column then generate the smaller mass squared difference 
$\Delta m^2_\mathrm{sol}:= m_2^2 - m_1^2 $ and can account for a large solar neutrino 
mixing $\theta_{12}$. Due to the small difference of the Majorana phases
corresponding to $m_1$ and $m_2$, the RG effects for $\theta_{12}$ 
in see-saw models are generically larger than for the other lepton mixing 
angles if $\theta_{23}$ is large already at high energy 
\cite{Antusch:2002hy,Antusch:2002fr}. 
Depending on $\tan \beta$ and the neutrino mass scale, 
they can be sizable or lead only to rather small corrections. 
For partially degenerate neutrinos with a mass scale up to 
about $0.15$ eV, the running  
in general does not cause problems with naturalness, but it allows for a 
wider range of possible mass and mixing patterns at high energy. 
A careful analysis has to include RG effects and we have provided 
figures where estimates for the corrections can easily be read off.

In the classes of type II see-saw models with sequential right-handed 
neutrino dominance for the type I 
contribution to the neutrino mass matrix and real vacuum alignment, 
the solar and the atmospheric neutrino mixings $\theta_{12}$ and $\theta_{23}$ 
are independent of the type II mass scale $m^{\mathrm{II}}$ and of the  
complex phases of the neutrino Yukawa matrix. This provides a natural way 
to upgrade these types of models continously from hierarchical neutrino 
mass spectra to 
partially degenerate ones, while maintaining the 
predictions for the two large lepton mixings. The mixing angle 
$\theta_{13}$ is generically small and furthermore decreases with 
increasing neutrino mass scale.  
In addition, we find that our scenario predicts that all observable CP phases, 
i.e.~the Dirac CP phase $\delta$ relevant for neutrino oscillations 
and the Majorana CP phases $\beta_2$ and $\beta_3$,  
become small as the neutrino mass scale increases. 
This implies in particular that the effective mass for 
neutrinoless double beta decay is approximately equal to the 
neutrino mass scale and therefore neutrinoless double beta decay 
will be observable if the neutrino mass spectrum is partially degenerate. 
 In our framework a partially degenerate neutrino mass spectrum is 
{\it a priori} as natural as a hierarchical spectrum. If neutrinos are 
partially degenerate, neutrinoless double beta decay has the potential to 
 measure the neutrino mass scale.

\section*{Acknowledgements}
We acknowledge support from the PPARC grant PPA/G/O/2002/00468.

\providecommand{\bysame}{\leavevmode\hbox to3em{\hrulefill}\thinspace}

\end{document}